%% file: ttbbj.tex
\documentclass[a4paper,11pt]{article}
\pdfoutput=1
\usepackage[english]{babel}
\usepackage{jheppub}
\usepackage{subfigure}
\usepackage{xspace}
\usepackage{amsmath,amssymb,textcomp,enumerate,alltt,xspace,multirow}
\usepackage{slashed}
\usepackage{multicol}

\usepackage{graphicx}
\usepackage{relsize}
\usepackage{pstricks}
\usepackage{color}
\usepackage{xcolor}

\input texmacros.tex

\preprint{
\begin{flushright}
PSI-PR-19-16\\
ZU-TH 38/19\\ 
\end{flushright}
}

\title{NLO QCD predictions for $\boldsymbol{\ttbar\bbbar}$ production in
association with a light jet at the LHC}

\author[a]{Federico Buccioni,}
\author[b]{Stefan Kallweit,}
\author[a]{Stefano Pozzorini}
\author[c]{and Max F. Zoller}
\emailAdd{buccioni@physik.uzh.ch}
\emailAdd{stefan.kallweit@cern.ch}
\emailAdd{pozzorin@physik.uzh.ch}
\emailAdd{max.zoller@psi.ch}

\affiliation[a]{
Physik-Institut, Universit{\"a}t Z{\"u}rich, Winterthurerstrasse 190, CH-8057 Z{\"u}rich, Switzerland
}
\affiliation[b]{
Dipartimento di Fisica, Universit\`a degli Studi di Milano-Bicocca and
INFN, Sezione di Milano-Bicocca, I--20126, Milan, Italy
}
\affiliation[c]{
Paul Scherrer Institut, CH-5232 Villigen PSI, Switzerland
}
\abstract{
Theoretical predictions for \ttbb production are of
crucial importance for $\ttbar H$ measurements in the $H\to \bbbar$ channel
at the LHC.
To address the large uncertainties associated with the 
modelling of extra QCD radiation in \ttbb events,
in this paper we present a calculation of $pp\to\ttbbj$
at NLO QCD.
The behaviour of NLO corrections is analysed in a variety of observables,
and to assess theoretical uncertainties we use factor-two rescalings as well
as different dynamic scales.
In this context, we propose a systematic alignment of dynamic scales that 
makes it possible to disentangle normalisation and shape uncertainties in a
transparent way.
Scale uncertainties at NLO are typically at the level of 20--30\% in
integrated cross sections, and below 10\% for the shapes of distributions.
The kinematics of QCD radiation is investigated in detail, including 
the effects of its recoil on the objects of the \ttbb system.
In particular, we discuss various
azimuthal correlations that allow one to characterise the
QCD recoil pattern in a precise and transparent way. 
In general, the calculation at hand provides   
a variety of precise benchmarks that can be used to validate the modelling of 
QCD radiation in \ttbb generators.
Moreover, as we will argue,  $pp\to \ttbbj$ at NLO entails information that
can be used to gain insights into the perturbative convergence of the
inclusive \ttbb cross section beyond NLO.
Based on this idea, we address the issue of the large NLO $K$-factor
observed in $\sigma_{\ttbb}$, and we provide evidence that supports 
the reduction of this $K$-factor through a mild adjustment of the
QCD scales that are conventionally used for this process.
The presented $2\to 5$ NLO calculations have been carried out 
using \OLtwo in combination with {\Sherpa} and {\sc Munich}.
}


\begin{document}
\maketitle

\flushbottom

\section{Introduction}

The associated production of top- and bottom-quark pairs at hadron colliders
is an especially interesting process.  From the theoretical point of view,
it offers rich opportunities to investigate the dynamics of QCD
in the presence of multiple scattering particles and energy scales.
In particular, higher-order calculations of $pp\to\ttbb$ raise non-trivial questions related to
the mass gap between $m_b$ and $m_t$, the choice of QCD scales, and 
the convergence of the perturbative expansion.
Further strong motivation for a deeper understanding of \ttbb production 
comes from its critical role as irreducible background 
to $\ttbar H$ production  with $H\to \bbbar$ 
at the LHC~\cite{Aaboud:2017rss,Sirunyan:2018mvw,CMS:2019lcn}.
In this context, the modelling of 
$pp\to\ttbb$ represents the main source of uncertainty in
$\ttbar H(\bbbar)$ measurements. Thus, improving the 
theoretical description of
the \ttbb background is of great importance 
for the sensitivity of $\ttbar H(\bbbar)$ analyses 
at the High-Luminosity
LHC~\cite{Cepeda:2019klc}.
Precise theoretical calculations for \ttbb production are relevant also for direct
experimental studies of this process,  and recent
measurements of the \ttbb cross section~\cite{Sirunyan:2017snr,Aaboud:2018eki,CMS:2019dij} 
tend to exceed theory predictions by \percentrange{30}{50}.


At leading order~(LO) in QCD, the \ttbb cross section is proportional to
$\as^4$ and suffers from huge scale uncertainties. 
Next-to-leading order (NLO) 
QCD calculations~\cite{Bredenstein:2009aj,Bevilacqua:2009zn,Bredenstein:2010rs}
reduce scale uncertainties to \percentrange{20}{30}, but the level of 
precision and the size of the corrections 
depend in a critical way on the choice of the renormalisation scale $\mur$.
In this respect, in order to avoid
an excessively large NLO $K$-factor, it was found that 
the value of $\mur$ should be chosen in the vicinity 
of the geometric average of the
energy scales of the \ttbar and 
\bbbar systems~\cite{Bredenstein:2010rs}.

Calculations of $pp\to \ttbb$ based on the five-flavour~(5F)
scheme~\cite{Bredenstein:2009aj,Bevilacqua:2009zn,Bredenstein:2010rs}, where
$b$-quarks are treated as massless partons, are applicable 
only to the phase space with two resolved $b$-jets,
while including $b$-mass effects in the four-flavour~(4F) scheme makes it
possible to obtain NLO predictions in the full $\ttbar+b$-jets phase
space~\cite{Cascioli:2013era}, including regions where one $b$-quark is 
unresolved.
The choice of the 4F scheme as opposed to the 5F scheme is also supported by
the fact that initial-state $g\to \bbbar$ splittings play a marginal
role in $\ttbar+b$-jets production, while the vast majority of $b$-jets 
originate via initial-state gluon radiation with subsequent $g\to
\bbbar$ splittings~\cite{Jezo:2018yaf}.


In order to be applicable to $\ttbar H (\bbbar)$ measurements, NLO
calculations of $pp\to\ttbb$ need to be matched to parton showers. 
Nowadays, this can be achieved within various Monte Carlo
frameworks~\cite{Garzelli:2014aba, Bevilacqua:2017cru,
Cascioli:2013era,
Alwall:2014hca, Jezo:2018yaf, Bellm:2015jjp}, using  
different matching methods and parton showers.
Some of these generators are in good mutual agreement, but the overall
spread of Monte Carlo predictions suggests that \ttbb modelling
uncertainties may significantly exceed the level of QCD scale variations,
thereby spoiling NLO accuracy~\cite{deFlorian:2016spz}.
In this context, the uncertainties related to the 
modelling of extra QCD radiation that accompanies \ttbb 
production play a dominant role.


Motivated by these observations, in this paper we present
a NLO QCD calculation of \ttbb production in association with one additional 
jet at the LHC.%
\footnote{Preliminary results of this project have been presented at
  QCD@LHC~2018~\cite{talkBuccioniQCDatLHC2018}
  and HP2~2018~\cite{talkBuccioniHP22018}.
}
Bottom-mass effects are included throughout using the 4F scheme.
For the calculation of the required $2\to 5$ one-loop amplitudes, 
which involve up to 25'000 diagrams in a single partonic channel,
we use the latest version of the \OpenLoops program~\cite{Buccioni:2019sur}, where
scattering amplitudes are computed with the new on-the-fly reduction method
presented in~\cite{Buccioni:2017yxi}. For the 
calculation of hadronic cross sections, \OLtwo is interfaced with 
\Sherpa~\cite{Krauss:2001iv,Gleisberg:2008ta,Gleisberg:2008fv,Gleisberg:2007md}
and, alternatively, with {\sc Munich}\footnote{\Munich{} is the 
abbreviation of ``MUlti-chaNnel Integrator at Swiss~(CH) precision'' --- an automated parton-level
NLO generator by S.~Kallweit.}.

We discuss NLO predictions for $pp\to \ttbbj$ at 13\,TeV with emphasis on
the assessment of perturbative uncertainties.  
To this end, we study conventional scale variations as well as different
dynamic scales, and we point out that the effects of these two kinds of
scale uncertainties are largely correlated.
Based on this observation, we propose the idea of aligning dynamic scales to
a natural scale, which can be defined using the maxima of the NLO variation
curves as a reference.
This prescription makes it possible to disentangle the effects of factor-two
variations and dynamic scale variations in a way that provides a more
transparent picture of normalisation and shape uncertainties.

To characterise the behaviour of QCD radiation in \ttbb events, we
consider kinematic distributions in the hardest light jet as well as recoil 
effects on the various objects of the \ttbb system.  To
this end, we introduce azimuthal angular correlations that provide a
transparent and perturbatively stable picture of recoil effects.  Our NLO
predictions for these and various other observables
can be used as precision benchmarks to validate the modelling of QCD radiation 
in \ttbb generators.

Finally, we exploit the calculation at hand to address the issue of the
large NLO $K$-factor observed in the integrated \ttbb cross
section~\cite{Jezo:2018yaf}.  In this respect, we note that 
the NLO corrections to $pp\to \ttbbj$ 
correspond to the same order in $\alphaS$ as
the NNLO corrections to inclusive \ttbb production, \ie $\ord(\alphaS^6)$. 
Thus they entail (partial) information on the behaviour of
$\sigma_{\ttbb}$ beyond NLO.
Based on this idea, we  use the \ttbbj cross section at NLO to
identify an optimal scale choice for the process $pp\to \ttbb$.
The results of this analysis support a slight adjustment of the
conventional \ttbb scale choice, which results in 
a reduction of the \ttbb $K$-factor and is also expected
to attenuate NLO matching uncertainties.


The paper is organised as follows. 
In \refses{se:ingredients}{se:technical}
we outline the main ingredients of $pp\to \ttbbj$ at NLO, and we document the
employed input parameters, scale choices 
and acceptance cuts.
In \refse{se:fidXS} we study the integrated cross sections and their scale
dependence, and we check the safeness of our predictions 
with respect to Sudakov logarithms beyond NLO.  
Moreover, we propose the idea of
disentangling shape and normalisation uncertainties by means of an alignment
prescription for dynamic scales.
Differential observables and shape uncertainties are presented in
\refse{se:ttbbjdist}, where we also discuss recoil
effects.
Finally, in \refse{se:ttbb_tuning} we use \ttbbj NLO predictions 
to identify an improved scale choice for inclusive \ttbb
production.
Our main findings are summarised in \refse{se:conclusions}.

\section{Ingredients of the calculation} 
\label{se:ingredients}

\renewcommand{\arraystretch}{1.3}
\begin{table}[t]
\vspace*{0.3ex}
\begin{center}
\begin{tabular}{cc|c|c|c}
order & type   & channel             & \# diagrams    & \# crossings $\times$ flavours
\\\hline
LO    & trees  & $gg\to\ttbb g$      &   393          &  $1\times 1$  \\
      &        & $q\bar q\to\ttbb g$ &   66           &  $6\times 4$  \\\hline
NLO   & loops  & $gg\to\ttbb g$      &  25431      &  $1\times 1$  \\ 
      &        & $q\bar q\to\ttbb g$ &  3534       &  $6\times 4$  \\\hline
NLO   & trees  & $gg\to\ttbb gg$             &  5190      &   $1\times 1$    \\
      &        & $q\bar q\to\ttbb gg$        &  795       &   $7\times 4$ \\
      &        & $q\bar q\to\ttbb q\bar q$   &  204       &   $4\times 4$  \\
      &        & $q\bar q\to\ttbb q'\bar q'$ &  102       &   $4\times 12$ \\
\end{tabular}
\end{center}
\caption{Independent partonic channels contributing to $pp\to \ttbb j$ at
NLO.
For each class of crossing-related processes we indicate 
a representative process, the number of colour-stripped diagrams, and
the number of crossings and quark-flavour assignments, $q,q'=u,d,c,s$, $q\neq q'$.
In \OL, each Feynman diagram corresponds to $3^{n_4}$ 
colour-stripped diagrams, where $n_4$ is the number of quartic gluon
vertices in the diagram at hand (typically $n_4=0$).
}
\label{tab:Ndiagrams}
\end{table}

\subsection[\ttbbj production in the 4F scheme]{$\boldsymbol{\ttbbj}$ production in the 4F scheme}

We investigate NLO QCD corrections to hadronic \ttbbj production in
the 4F scheme, \ie we treat not only top quarks, but also bottom quarks
with a finite mass throughout. The non-vanishing bottom mass renders $g\to b\bar{b}$ splittings finite,
which allows us to investigate also observables with unresolved $b$-jets and to apply the
experimentally favoured definition of $b$-jets as all hadronic jets that contain at
least one bottom \mbox{(anti-)quark} at the parton level. In particular, jets resulting from the clustering
of $b$ and $\bar{b}$ partons are considered $b$-jets as well. Accordingly, only hadronic jets
that are constituted from light quarks $q=d,u,s,c$ and gluons are considered light jets.
In the 4F scheme, since no bottom (anti-)quarks appear as proton
constituents, no further bottom (anti-)quarks are generated at NLO QCD. Thus all $b$-jets
are generated by Feynman diagrams that contain exactly one $\bbbar$ pair.
Input parameters, renormalization scheme and parton-distribution functions~(PDFs) are chosen
according to the 4F scheme, as detailed in \refse{se:input}.

\begin{figure}[t]
\centering
\plotnlodiag{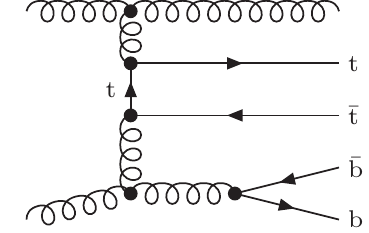}
\plotnlodiag{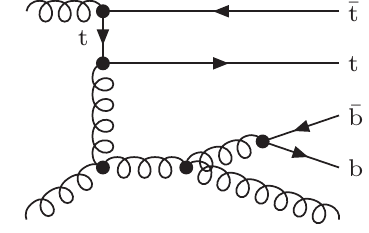}
\plotnlodiag{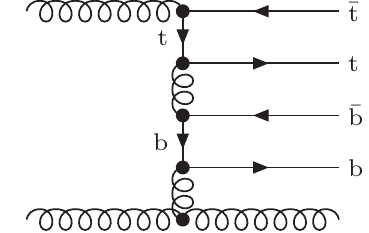}\\[5mm]
\caption{Selected Born diagrams in the $gg\to \ttbb g$ channel.
\vspace*{2ex}
}
\label{fig:diagramsborn}
\end{figure}
The independent partonic channels contributing to $pp\to \ttbbj$ at NLO
are summarised in \refta{tab:Ndiagrams}
together with the number of Feynman diagrams and   
crossing/flavour symmetries.
At LO, \ttbbj production involves the two crossing-independent channels
$gg\to t\bar{t}b\bar{b}g$ and
$q\bar{q}\to t\bar{t}b\bar{b}g$
with \mbox{$q=d,u,s,c$}, where the latter 
gives rise to six quark--anti-quark and gluon--(anti)quark channels 
via permutations of $q,\bar q, g$.

\reffi{fig:diagramsborn} illustrates sample diagrams for the 
gluon--gluon channel, which is by far the dominant channel, 
with a contribution of
about $77\%$ ($qg$: $21\%$, $q\bar{q}$: $2\%$). 
The dominant $gg\to \ttbb g$ topologies are those where the \bbbar{} pair is emitted from a $g\to\bbbar$ splitting
and the final-state gluon results from an initial-state $g\to gg$ splitting, while the \ttbar pair
is produced in a $t$-channel configuration. However, the impact of other topologies becomes
quite prominent in certain phase-space regions, like \eg at high invariant mass
or $\Delta R$ separation of the
\bbbar{} system. See also \reffi{fig:ttbbtopologies} for the dominant $gg\to
\ttbb$ topologies.

At NLO in QCD, as usually the process receives contributions both from virtual and real
corrections, which are separately divergent. To mediate these divergences between the different
phase spaces, we rely on the dipole-subtraction formalism~\cite{Catani:1996vz}
in its extension to massive QCD partons~\cite{Catani:2002hc}.

\begin{figure}[t]
\centering
\plotnlodiag{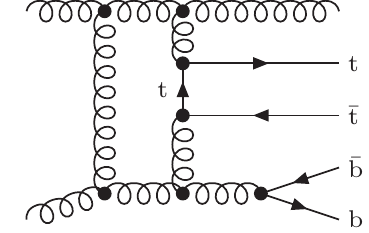}
\plotnlodiag{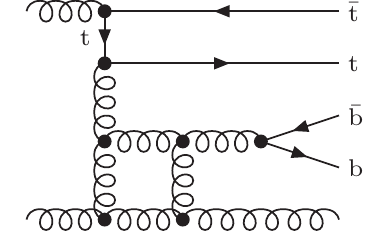}
\plotnlodiag{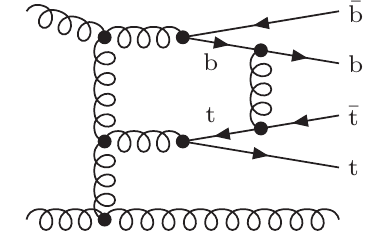}
\\[10mm]
\plotnlodiag{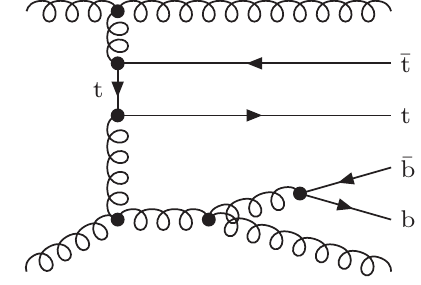}
\plotnlodiag{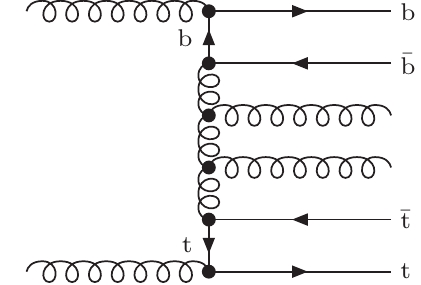}
\plotnlodiag{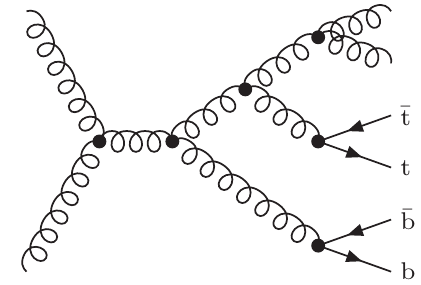}
\\[5mm]
\caption{Selected $gg\to \ttbb g$ one-loop diagrams
(first row) and $gg\to \ttbb gg$  real-emission
diagrams (second row).}
\vspace*{2ex}
\label{fig:diagramsnlo}
\end{figure}
The virtual corrections are constituted from both diagrams with a closed quark loop and diagrams that are
generated from the LO ones by exchanging a virtual gluon between any of the external or internal legs.
Since all involved partons interact under QCD, the number of loop diagrams is more than
a factor of 50 larger than the number of Born diagrams in the respective channels (see \refta{tab:Ndiagrams}).
While the quark-loop diagrams contain up to pentagon functions,
the gluon-exchange diagrams require up to heptagon functions.
Some sample diagrams for the latter are shown in \reffi{fig:diagramsnlo} (first row),
again for the dominant $gg$ channel only.

The real-correction channels are constructed from the LO ones by either emission of another gluon
or by the splitting of a gluon into a light $q\bar{q}$ pair. Including crossings of light partons
between initial and final states, the channels listed in \refta{tab:Ndiagrams} result.
In \reffi{fig:diagramsnlo} (second row)
we depict sample diagrams for the dominant all-gluon channel.

\subsection{Tools and validation}
\label{se:setup:tools}

The calculations presented in this paper have been performed with the 
automated frameworks \SherpaOpenLoops and \MunichOpenLoops. Each of them completes
the full chain of operations --- from process definition to collider 
observables --- that enter NLO QCD simulations at parton level. 

In both frameworks virtual amplitudes are provided by
\OpenLoops{}~2~\cite{Buccioni:2019sur}, the latest version of the \OpenLoops{}
matrix-element generator. One of the of main novelties of \OpenLoops{}~2,
which is used for the first time in the calculation at hand, 
is the combination of the original open-loop
algorithm~\cite{Cascioli:2011va} with the recently proposed
on-the-fly reduction method~\cite{Buccioni:2017yxi}.
In this approach, the construction of
loop amplitudes and their reduction to scalar integrals are combined 
in a single
numerical recursion, which makes it possible to generate
one-loop amplitudes in a way that avoids high tensorial ranks at all stages of the
calculations.
This results in a significant speed-up for multi-leg processes.
Specifically, for the process at hand, the excellent CPU performance
of \OpenLoops{}~1 is further improved by a factor of three.
For the treatment of numerical instabilities, the on-the-fly reduction
algorithm is equipped by an automated stability system that combines
analytic expansions together with a novel hybrid-precision system.
The latter detects residual instabilities based on the analytic 
structure of reduction identities and cures them by switching 
from double (dp) to quadruple (qp) precision. Thanks to the local and highly targeted usage of
qp, the typical qp overhead wrt dp
evaluation timings is reduced from two orders of
magnitude to a few percent.

The only external ingredients required by \OpenLoops{}~2 
are the scalar integrals~\cite{Denner:2010tr}, which
are provided by the \Collier library~\cite{Denner:2014gla,Denner:2016kdg} by default, 
or by the {\sc OneLOop} library~\cite{vanHameren:2010cp} for exceptional qp
evaluations.
All amplitudes have been thoroughly validated against
\OpenLoops{}~1~\cite{Cascioli:2011va}, where the reduction is carried 
out based on the Denner--Dittmaier techniques~\cite{Denner:2002ii,Denner:2005nn}
available in \Collier or, alternatively, using {\sc CutTools}~\cite{Ossola:2007ax}, which 
implements the OPP method~\cite{Ossola:2006us}, together with the {\sc OneLOop} 
library~\cite{vanHameren:2010cp} for scalar integrals.
Additionally, matrix elements have been cross-checked against the completely independent
generator \Recola~\cite{Actis:2016mpe,Denner:2017wsf}.

All remaining tasks, \ie the bookkeeping of
partonic subprocesses, phase-space integration, and the subtraction of QCD
bremsstrahlung, are supported by the two independent and fully
automated Monte Carlo generators, \Munich and \Sherpa.  

In \Sherpa, tree amplitudes are computed using \Comix~\cite{Gleisberg:2008fv},
a matrix-element generator based on the colour-dressed Berends-Giele recursive
relations~\cite{Duhr:2006iq}, while one-loop amplitudes are provided by \OpenLoops{}.
Infrared singularities are cancelled using the  
dipole subtraction method~\cite{Catani:1996vz,Catani:2002hc}, as automated in \Comix,
with the exception of K- and P-operators that are taken from the implementation  
described in~\cite{Gleisberg:2007md}. \Comix is also used for the evaluation
of all phase-space integrals. Analyses are performed with the help of
\Rivet~\cite{Buckley:2010ar}, which involves
the \FastJet{} package~\cite{Cacciari:2005hq,Cacciari:2011ma} to cluster partons into
jets.

The parton-level generator \Munich{} has been applied to several multi-leg processes
at NLO QCD and EW accuracy,
and as a key ingredient of the \Matrix{} framework~\cite{Grazzini:2017mhc} it has been
intensively applied to boson and diboson production at NNLO QCD.
\Munich{} provides a very efficient multi-channel phase-space integration with several
optimizations for higher-order applications.
All tree-level and one-loop amplitudes are supplied by \OpenLoops{} through a fully automated
interface.
The implementation of the massive dipole subtraction
formalism used in the present calculation has been extensively tested in
the context of off-shell top-pair production in the 4F scheme~\cite{Cascioli:2013wga},
and very recently in the NNLO QCD production of \ttbar pairs~\cite{Catani:2019iny,Catani:2019hip}.
The implementation of phase-space cuts at generation and analysis level, as well as the event
selection including jet algorithms are realized directly in \Munich{}, without relying on external
tools. Also the calculation of arbitrary (multi-)differential observables and the setting of
dynamic scales are handled internally. Thereby \Munich{} provides an independent
cross-check of basically all remaining steps of the working chain.

Both tools have been validated extensively against each other for a representative selection
of the results presented in this paper. All cross sections binned in $b$-jet and light-jet
multiplicities (see \reftastwo{tab:XSA}{tab:XS}) have been validated at a precision level of $0.3\%$ throughout for all scale choices. Moreover, most of the differential distributions presented in
\refse{se:ttbbjdist} have been cross-checked at the NLO level.
For all compared observables we find agreement on the level expected from the statistical
uncertainties of the two independent calculations.

\section{Technical aspects and setup}
\label{se:technical}

In this section we specify the input parameters, PDFs, scale choices 
and acceptance cuts used in the 
calculations presented in \refses{se:fidXS}{se:ttbb_tuning}.

\subsection{Input parameters, PDFs and scale choices}
\label{se:input}

Heavy-quark mass effects are included throughout using
\be
 m_{t} = 172.5 \;\GeV\,,\qquad     m_{b} = 4.75 \;\GeV\,.
\ee
All other quarks are treated as massless in the perturbative part of the
calculations.  Since we use massive $b$-quarks, for the PDF evolution and
the running of $\as$ we adopt the 4F scheme.
Thus, for consistency, we renormalise $\as$ in the decoupling scheme, where
top- and bottom-quark loops are subtracted at zero momentum transfer.  In
this way, heavy-quark loop contributions to the evolution of the strong coupling
are effectively described at first order in $\as$ through the virtual corrections.

We present predictions for $pp\to \ttbbj$ 
at $\sqrt{s}=13$\,TeV.
At LO and NLO
we use throughout the 4F {\tt NNPDF} parton
distributions~\cite{Ball:2014uwa} at NLO, 
and the corresponding strong coupling.%
\footnote{More precisely we use the
{\tt NNPDF30\_nlo\_as\_0118\_nf\_4} parton distributions, 
as implemented in LHAPDF~\cite{Buckley:2014ana},
where $\asfour(M_Z)=0.112$, which corresponds to $\asfive(M_Z)=0.118$.
}
PDF uncertainties are expected to play a rather
subleading role, similarly as for $pp\to \ttbb$~\cite{Jezo:2018yaf}.  Thus we will base our
predictions on the nominal PDF set, restricting our assessment of
theoretical uncertainties to perturbative scale variations.%
\footnote{Using 100 replicas of the PDF set at hand
we have checked that PDF uncertainties are at the level of 
10\% for the integrated $\ttbbj$ cross section and grow slowly with the 
$\pT$ of the various final-state objects, reaching at
most 20\% in the regions 
where event rates are suppressed by 
two orders of magnitude.
We refrain from reporting further details on PDF uncertainties since they 
are strongly correlated to the ones observed in inclusive $\ttbb$ production~\cite{Jezo:2018yaf}
and thus only marginally relevant for the theoretical questions addressed in 
this paper.}

\subsection{Renormalisation and factorisation scales}
\label{se:scalechoice}

Since it scales with $\as^5$, the \ttbbj cross section is highly sensitive
to the choice of the renormalisation scale $\mur$, and this choice 
plays a critical role for the stability of perturbative predictions.
Along the lines of~\cite{Cascioli:2013era,deFlorian:2016spz,Jezo:2018yaf}, we adopt a
dynamic scale that accounts for the 
fact that \ttbb production is characterised by two 
widely separated scales,
which are related to the \ttbar and \bbbar systems.
To this end we define
\begin{equation}
\label{eq:mubbtt}
\mu^2_{\bbbar}=E_{\rT,b}E_{\rT,\bar b},
\qquad
\mu^2_{\ttbar}=E_{\rT,t}E_{\rT,\bar t},
\qquad
m^2_{\bbbar}=(p_b+p_{\bar b})^2\,,
\end{equation}
where the transverse energies $E_{\rT,i}=\sqrt{m_i^2+p^2_{\rT,i}}$ are defined in terms of 
the rest masses $m_i$ and the transverse momenta $p_{\rT,i}$ 
of the bare heavy quarks, without applying any jet algorithm at NLO. 
Also ${m^2_{\bbbar}}$ is defined in terms of the bare four-momenta of the
(anti-)$b$ quarks.
As default choice for the renormalisation scale we adopt
the geometric average of the various transverse energies and momenta of the
\ttbbj system, 
\bea
\label{eq:murdef}
\mudef\,(\xir) &=&
\xir\, \mudef
\,=\, \xir  \Big(\mu^2_{\ttbar}\,\mu^2_{\bbbar}\,p_{\rT,j}\Big)^{1/5},
\eea
where the rescaling factor $\xir$ is typically varied in the 
range $[0.5,2]$.
This choice represents the natural generalisation of the
widely used scale~\cite{Cascioli:2013era, deFlorian:2016spz, Jezo:2018yaf}
\bea
\muttbb\,(\xir) &=&
\xir\, \muttbb \,=\, \xir  
\left(\mu_{\ttbar}\,\mu_{\bbbar}\right)^{1/2}
\label{eq:muttbb}
\eea
for \ttbb production.%
\footnote{The choices  \refeq{eq:murdef}--\refeq{eq:muttbb} are motivated by
the fact that, to lowest order in the strong coupling,
$\alphaS^5(\muttbbj)=\alphaS^2(\mu_{\ttbar})\,\alphaS^2(\mu_{\bbbar})\,\alphaS(p_{\rT,j})$
and $\alphaS^4(\muttbb)=\alphaS^2(\mu_{\ttbar})\,\alphaS^2(\mu_{\bbbar})$.  In
this way, the coupling factors associated with the production of the
$\ttbar$ and $\bbbar$ systems, plus the additional light jet for \ttbbj
production,
are effectively
evaluated at the corresponding characteristic scales, $\mu_{\ttbar}$,
$\mu_{\bbbar}$ and $p_{\rT,j}$, avoiding large
logarithms associated with the evolution of $\alphaS$.}
The additional light-jet $p_\rT$ that enters~\refeq{eq:murdef} is defined
using an auxiliary\footnote{For the definition of physical observables a
conventional anti-$k_\rT$ algorithm is used (see below).} $k_{\rT}$-jet algorithm with
$R = 0.4$, which is applied only to massless partons, \ie excluding top and
bottom quarks from the recombination, and is free from any restriction in
$\pT$ and rapidity.

\begin{figure}[t!]
\centering
\subfigure[]{
\includegraphics[width=0.25\textwidth]{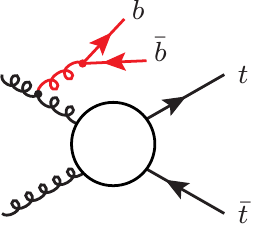}
\label{subfig:ISdom}
}\hspace{.15\textwidth}
\subfigure[]{
\includegraphics[width=0.3\textwidth]{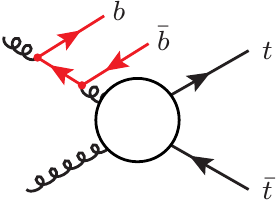}
\label{subfig:FSdom}
}
\caption{Generic leading-order $gg\to \ttbb$ topologies
with final-state (a) and initial-state (b) 
$g\to \bbbar$ splittings. The bulk of the \ttbb cross section is dominated
by topologies of type (a) with rather collinear splittings, while 
initial-state collinear splittings become important in the 
region of large $\DRbb$~\cite{Jezo:2018yaf}. 
}
\vspace*{2ex}
\label{fig:ttbbtopologies}
\end{figure}

In order to assess shape uncertainties, we consider 
three alternative dynamic scales with different kinematic dependences.
The first one is defined as
\bea
\label{eq:murmbb}
\mumbb(\xir) \,=\, 
\xir\, \mumbb \,=\, 
\left(\mu^2_{\ttbar}\,m_{\bbbar}\,E_{\rT,\bbbar}\,p_{\rT,j}\right)^{1/5},
\eea
where the $\bbbar$ system enters through its invariant mass 
and its total transverse energy,
$E_{\rT,\bbbar}=\sqrt{m_{\bbbar}^2+(\vec p_{\rT,b}+\vec p_{\rT,\bar b})^2}$. 
This choice is motivated by the fact that $m_{\bbbar}$ and $E_{\rT,\bbbar}$
correspond to the virtualities of the QCD branching processes that
dominate \ttbb production, namely initial-state $g\to gg$ splittings
followed by a final-state $g\to \bbbar$ splittings (see
\reffi{fig:ttbbtopologies}).

As further alternatives we consider two other dynamic scales,
\bea
\label{eq:murht}
\muht\,(\xir) &=& \xir\, \muht =\xir \frac{H_\rT}{5},
\eea
and
\beqar
\label{eq:murhbb}
\muhtbbj\,(\xir) &=& \xir\, \muhtbbj \,=\, \xir  
\left(\mu_{\ttbar}\,\frac{\htjets}{3}\right)^{1/2}\,,
\eeqar
which are defined
in terms of the transverse energies of the jets,
\begin{equation}
\htjets
=  
\sum_{i=b, \bar b, g,q,\bar q} E_{\rT,i}\,,
\end{equation}
and the total transverse energy,
\begin{equation}
\HT
=  
\htjets+\sum_{i=t,\bar t} E_{\rT,i}\,.
\label{eq:htdef}
\end{equation}
Here $E_{\rT,j} = p_{\rT,j}$ for
massless partons, and the sums run over all final-state QCD partons, always including NLO
radiation and excluding only top quarks in the case of $\htjets$.

The factorisation scale  $\muf$ represents the maximum transverse momentum
for initial-state radiation that is resummed in the PDFs. Thus it is typically chosen of
the order of the halved hard-scattering energy.
Following~\cite{Cascioli:2013era, Jezo:2018yaf} 
we use\footnote{To be precise, the choice 
\refeq{eq:muf} 
agrees with the one used in~\cite{Jezo:2018yaf} but differs from 
the choice $\muf=\frac{1}{2}\sum_{i=t, \bar t} E_{T,i}$
made in~\cite{Cascioli:2013era}. 
However, this difference has a minor impact on 
our predictions.}
\begin{equation}
\label{eq:muf}
\muf=\xif\,\frac{H_\rT}{2}\;,
\end{equation}
where $\xif\in [0.5,2]$.

Our nominal predictions correspond to $\xir=\xif=1$, and to quantify scale
uncertainties we take the envelope of the seven-point variation
$(\xir,\xif)=(0.5,0.5)$, $(0.5,1)$, $(1,0.5)$, $(1,1)$, $(1,2)$, $(2,1)$, $(2,2)$.

\subsection{Jet observables and acceptance cuts}
\label{se:cuts}

\renewcommand{\arraystretch}{1.3}
\begin{table}[t]
\begin{center}
\begin{tabular}{c|cc|cc|cc}
region    & \regionttb  & \regionttbb & \regionttbj & \regionttbbj   & \regionttbjj   & \regionttbbjj \\\hline
$\Nbmin$ &    1 &    2 &    1  &    2   &    1   &    2 \\
$\Njmin$ &    0 &    0 &    1  &      1 &    2   &    2 \\ 
\end{tabular}
\end{center}
\caption{Naming scheme for phase-space regions with different inclusive 
multiplicities of $b$-jets ($\Nb\ge \Nbmin$) and light jets ($\Nj\ge \Njmin$) 
that pass the acceptance cuts \refeq{eq:jetcuts}.
}
\vspace*{1ex}
\label{tab:ttbblabelling}
\end{table}

For the reconstruction of jets we use the anti-$k_\rT$~\cite{Cacciari:2008gp}
algorithm with $R=0.4$. We select $b$-jets and light jets that fulfil
the acceptance cuts
\bea
\label{eq:jetcuts}
\pT>\pTcut&=& 50\,\GeV,\qquad
|\eta|<2.5\,.
\eea
We define as $b$-jet a jet that contains at least one $b$-quark, \ie jets that
contain a $\bbbar$ pair arising from a collinear $g\to \bbbar$ splitting are
also tagged as $b$-jets.\footnote{This prescription corresponds to 
a realistic experimental $b$-tagging, in the sense that the 
presence of one (or more) $b$-partons is sufficient to tag a jet as a $b$-jet.
In this respect we note that
jets containing a $g\to \bbbar$ splitting cannot be resolved in the 5F
scheme, since they would lead to 
uncancelled collinear singularities.
For this reason, in the 5F scheme an unphysical $b$-tagging prescription 
is used according to which 
jets containing a 
$g\to \bbbar$ splitting are regarded as light jets.
}
Top quarks are kept stable throughout.
When studying \ttbbj production, we 
categorise events
according to the number $\Nb$ of $b$-jets 
and the number $\Nj$ of light jets that fulfil the acceptance
cuts~\refeq{eq:jetcuts}.
We always consider inclusive phase-space regions with $\Nb\ge \Nbmin$ and 
$\Nj\ge \Njmin$, and we label them as indicated in
\refta{tab:ttbblabelling}. 
For the analysis of cross sections and distributions,
we always require one additional jet, and we 
consider an inclusive \regionttbj selection ($\Nbmin=1$)
and a more exclusive \regionttbbj selection  ($\Nbmin=2$).

\section{Integrated cross sections for $\boldsymbol{pp\to \ttbbj}$ at 13\,TeV}
\label{se:fidXS}

In this section we present numerical predictions for $pp\to \ttbbj$ at
$\sqrt{s}=13$\,TeV in the 4F scheme.  The results have been
obtained with \SherpaOpenLoops and \MunichOpenLoops, using the setup of \refse{se:technical}.
Top quarks are kept stable throughout, and we study cross sections and distributions
in the inclusive \regionttbj and \regionttbbj phase-space regions as defined in 
\refta{tab:ttbblabelling}, applying the acceptance cuts \refeq{eq:jetcuts}.
Perturbative scale uncertainties are assessed by means of seven-point 
factor-two scale variations and by comparing the various dynamic scales defined
in~\refse{se:scalechoice}.

\subsection{Renormalisation scale dependence}
\label{se:murscans}

\begin{figure}[t!]
\centering
\includegraphics[scale=0.75]{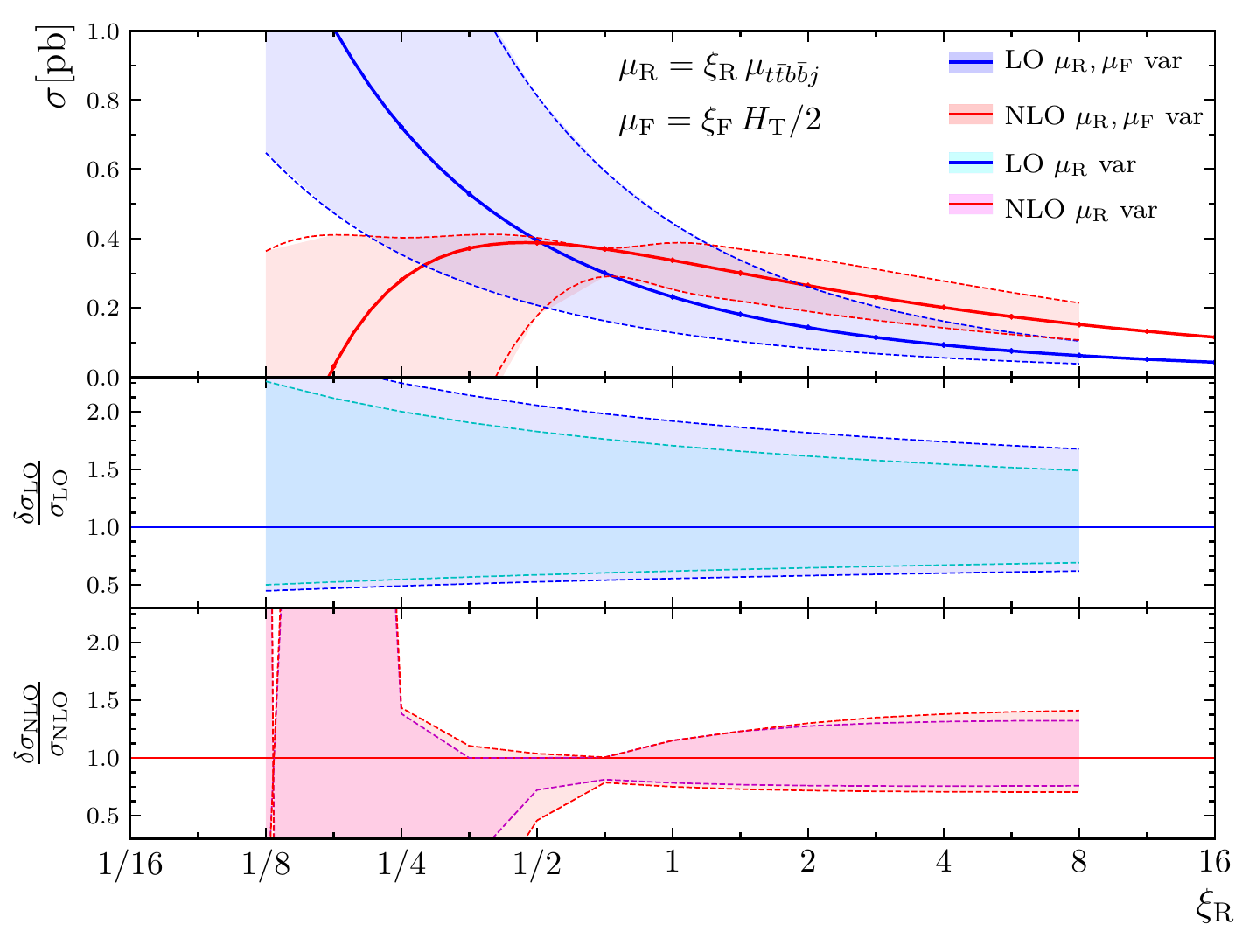}
\vspace*{-2ex}
\caption{Cross sections at $\sqrt{s}=13\,$TeV in the \regionttbbj phase
space. Predictions at LO (blue) and NLO
(red) with scales
$\mur=\xir \mudef$ and $\muF=\HT/2$ are plotted as a function 
of the renormalisation scale factor $\xir$.
The main frame presents absolute predictions and 
corresponding 7-point factor-two variations of $\muR$
and $\muF$, which are shown as uncertainty bands.
The relative impact of such variations at LO and NLO 
is displayed in the two ratio plots, which show 
also a second uncertainty band 
corresponding to pure factor-two variations of $\muR$
at fixed $\muF=\HT/2$.
}
\vspace*{2ex}
\label{fig:scannominal}
\end{figure}

A first picture of the perturbative behaviour of the \ttbbj cross section is 
displayed in \reffi{fig:scannominal}, where LO and NLO predictions 
based on the nominal scale choice \refeq{eq:murdef}
are plotted as a function of the renormalisation scale $\mur$.
For each value of $\muR$, the effect of factor-two scale variations is 
illustrated through two bands, which correspond to the 
variation of $\muR$ alone and the 
full 7-point variation of $\muR$ and $\muF$.
The results demonstrate that $\muF$ variations play only a marginal role, 
especially at NLO. Thus, in the following we will focus on the 
$\muR$ dependence.

\begin{figure}[t!]
\centering
\includegraphics[scale=0.75]{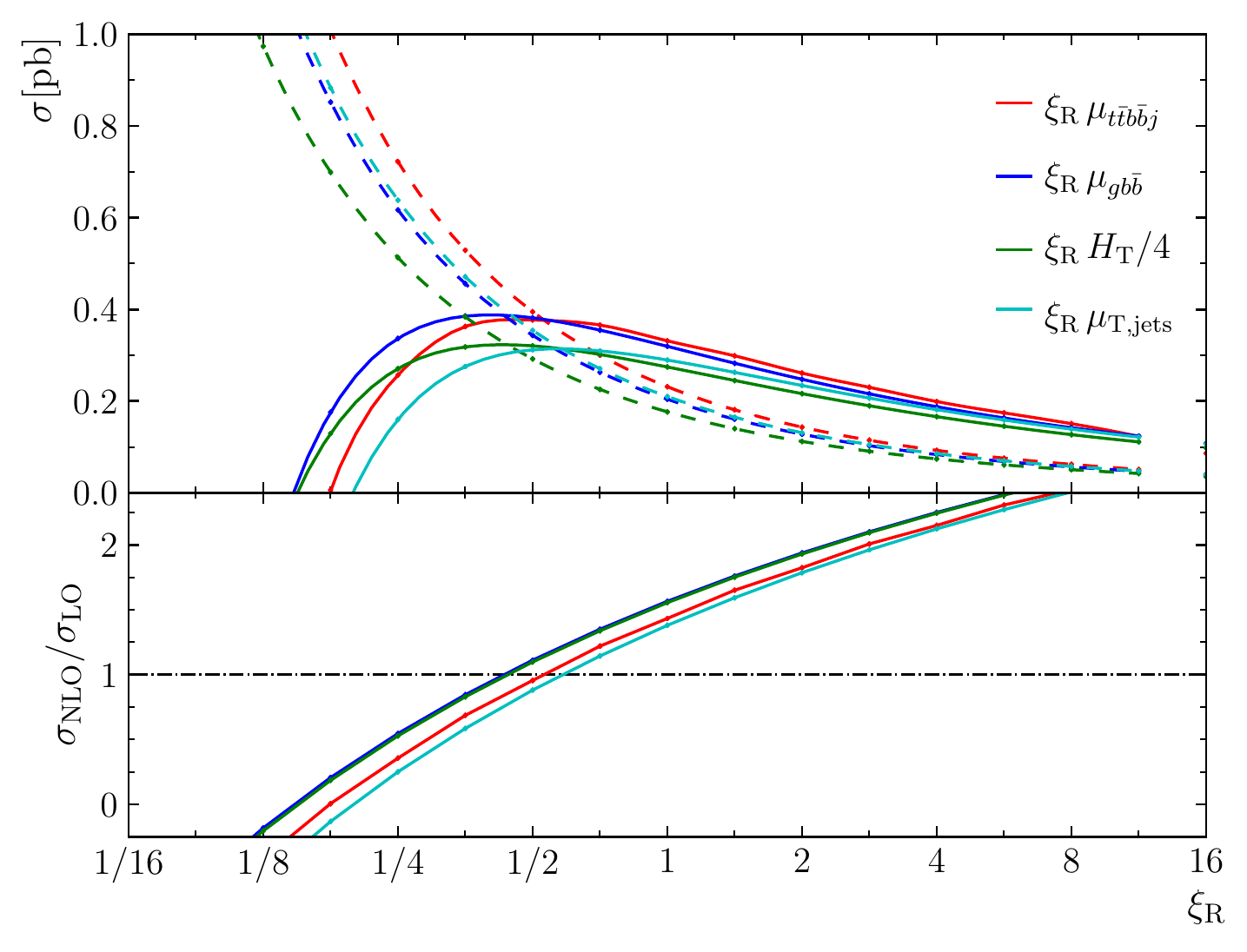}
\vspace*{-2ex}
\caption{Renormalisation-scale dependence of the LO (dashed) and NLO
(dotted) cross sections at $\sqrt{s}=13\,$TeV in the \regionttbbj phase space.  The different curves
correspond to the four dynamic scales defined in
\refeq{eq:murdef},\refeq{eq:murmbb}--\refeq{eq:murhbb}.  More precisely, instead of
$\muht=\HT/5$, the scale \HT/4 is shown in this plot.  Each scale is varied
around its nominal value ($\xir=1$) by a factor $\xir\in[1/16, 16]$, and
the factorisation scale is kept fixed at $\muF=\HT/2$.
Absolute predictions are shown in the main frame, while the 
ratio plot shows the NLO correction factor
$K(\xir)=\sigmaNLO(\xir)/\sigmaLO(\xir)$.
}
\vspace*{2ex}
\label{fig:scanNOTaligned}
\end{figure}

In \reffi{fig:scanNOTaligned} we plot the LO and NLO 
\ttbbj cross section as a function of $\mur$ for all four
dynamical scales defined in
\refeq{eq:murdef},\refeq{eq:murmbb}--\refeq{eq:murhbb}.
For each choice the renormalisation scale is 
varied around its nominal value by a factor 
$\xir\in[1/16, 16]$, while the factorisation scale is kept fixed at 
$\muf=\HT/2$.
The behaviour of the LO curves in \reffi{fig:scanNOTaligned} reflects 
the $\alphaS$-dependence of the LO cross section,  $\sigma_\LO\propto
\alphaS^5$, and corresponds essentially to the
running of $\alphaS$ to
the fifth power.  
To discuss the qualitative behaviour of \reffi{fig:scanNOTaligned} in more
detail, let us 
consider the effect of $\mur\to \xi \mur$ rescalings at LO,
\bea
\alphaS(\xi\,\mur)
&=&
\alphaS(\mur) \big[1+a_0(\mur)\ln\xi \big]^{-1}\,.
\label{eq:locrescalingA}
\eea
Here $a_0(\mur)=b_0 \alphaS(\mur)/(2\pi)=\ln^{-1}(\mur/\Lambda_\QCD)$, and
for small variations $\delta \xi$, %
\bea
\frac{\delta \alphaS^5}{\alphaS^5}
= -5 a_0(\mu)\, \frac{\delta \xi}{\xi}\,.
\label{eq:LOmurdep}
\eea
This is consistent with the LO curves of \reffi{fig:scanNOTaligned}, where
we observe that around the nominal scales ($\xi = 1$), reducing $\mur$ by a
factor 2 augments the LO cross sections by a factor close to 2 and vice
versa, which corresponds to $5 a_0(\mu)\sim 1$.

At NLO, the one-loop $\alphaS$-counterterm cancels the $\xi$-dependence at
$\ord(\alphaS\ln\xi)$, resulting in a 
significant reduction of scale variations.
In the vicinity of the nominal scales, factor-two
variations go down to 10--25\%,
depending on the type of scale and the direction of the variation.
As usually, the various NLO curves feature a stable point, which is 
located between $\xir=1/2$ and $1/3$. In the region below the maximum, 
the NLO curves start falling quite fast, and between $\xir=$1/6 and $1/8$ 
they lead to  negative cross sections.
To avoid such a pathologic perturbative behaviour,
the normalisation factors in the definition of $\muht$ and
$\muhtbbj$ have been chosen in such a way that factor-two variations of the
nominal scales do not enter the region below the NLO maximum.
Concerning the NLO correction factors, $K=\sigma_\NLO/\sigma_\LO$, at
$\xir\simeq 1$ we find $K\sim 1.5$ while the $K$-factor approaches one 
in the vicinity of the NLO maxima of the respective curves.

A striking feature of~\reffi{fig:scanNOTaligned} is that, in spite of the
rather different kinematic dependence of the various dynamic scales, the
observed LO and NLO scale variations and $K$-factors
have a fairly similar shape. 
In order to gain more insights into the origin of this behaviour, in the 
following we focus on the $\alphaS$-dependence of the LO cross section. For the 
differential and integrated cross sections let us define
\bea
\rd\sigma_\LO\big(\mudyn(\Phi)\big)=\alphaS^5\big(\mudyn(\Phi)\big)\,\rd\hat\sigma_\LO,\qquad
\hat \sigma_\LO = \int \rd\hat\sigma_\LO\,=\,
\int\rd \Phi\,
\,\frac{\rd\hat\sigma_\LO}{\rd\Phi}\,,
\label{eq:alphadepdef}
\eea
where $\mudyn(\Phi)$ is a certain dynamic scale,  $\Phi$ stands for the
fully-differential final-state phase space, and the convolution with PDFs as
well as acceptance cuts are implicitly understood.  For the integrated cross
section with dynamic scale $\mudyn$ we can write
\bea
\sigma_\LO(\mu_\dyn) &=& \int {\rd\hat\sigma_\LO}
\,\alphaS^5\big(\mu_\dyn(\Phi)\big)
\,=
\,\alphaS^5\left(\barmudyn\right)
\hat\sigma_\LO
\,,
\label{eq:averagedynscale}
\eea
where the result is expressed in terms of the $\alphaS$-free cross section
$\hat\sigma_\LO$ and the coupling factor $\alphaS^5(\barmudyn)$, which corresponds to the
average of $\alphaS^5\left(\mu_\dyn(\Phi)\right)$.
The above identity is nothing but a definition of the ``average'' scale
$\barmudyn$, which depends both on the functional form of $\mu_\dyn(\Phi)$
and on the applied phase-space cuts.
Let us now consider scale variations,
\bea
\sigma_\LO(\xi\,\mudyn) &=&
\int {\rd\hat\sigma_\LO}
\,\alphaS^5\big(\xi\,\mu_\dyn(\Phi)\big)\,.
\label{eq:rescaledLOdef}
\eea
The effect of $\mudyn\to \xi\,\mudyn$ on $\alphaS\big(\mudyn(\Phi)\big)$ can
be expressed as
\bea
\alphaS\big(\xi\,\mudyn(\Phi)\big) 
&=&
\alphaS(\xi\,\barmudyn) \left[
1+a_0(\xi\,\barmudyn)\ln\left(\frac{\xi\,\mudyn(\Phi)}{\xi\,\barmudyn}\right)
\right]^{-1}\nonumber\\
&=&
\alphaS(\xi\,\barmudyn) \sum_{n=0}^\infty\left[-
a_0(\xi\,\barmudyn)\ln\left(\frac{\mudyn(\Phi)}{\barmudyn}\right)
\right]^n\,,
\label{eq:locrescaling}
\eea
where the $\alphaS(\xi\,\barmudyn)$ prefactor on the rhs corresponds to a
trivial rescaling of $\barmudyn$, while the term between square brackets
depends on all moments of the distribution in
$\ln\left(\mudyn(\Phi)\right)$,
\bea
\big\langle\ln^n(\mudyn)\big\rangle &=& \frac{1}{\hat\sigma_\LO}\int \rd\hat\sigma_\LO
\,\left[\ln\big(\mu_\dyn(\Phi)\big)\right]^n\,.
\eea
Such moments may influence the scale dependence in a non-trivial way. 
However, their actual impact on the integrated cross section turns out to be
marginal.  
This is due to the fact that QCD cross sections are typically dominated by
phase-space regions with well-defined energy scales in the vicinity of the
thresholds for producing massive final states and passing acceptance cuts.
As a consequence, the distribution in $\ln\left(\mudyn(\Phi)\right)$ is
confined in the vicinity of its average value, $\ln(\barmudyn)$, and its
higher moments are rather strongly suppressed.  This implies
\bea
\big\langle\ln^n(\mudyn)\big\rangle &\simeq& \ln^n(\barmudyn)\,,
\eea
for all $n\ge 1$. More precisely, let us assume\footnote{For the process at
hand we have checked that, at LO, in the \regionttbbj phase space
the moments \refeq{eq:moments} are suppressed as
$X_n = \ord({10^{-n}})$ for $n=2,3,4,5,6$.
}
that
\bea
X_n &=& a^n_0(\xi\,\barmudyn) \big\langle 
\big(\ln(\mudyn) - \ln(\barmudyn)\big)^n\big\rangle \ll 1\,,
\label{eq:moments}
\eea
for $n\ge 2$. This implies that the expectation value of the rhs of
\refeq{eq:locrescaling} is dominated by the $n=0$ term.  Thus, under the
above assumptions, the scale dependence of the LO cross sections~\refeq{eq:rescaledLOdef}
can be approximated as
\bea
\sigma_\LO(\xi\,\mu_\dyn) &\simeq& 
\,\alphaS^5\left(\xi\,\barmudyn\right)
\hat\sigma_\LO
\,,
\label{eq:averagedynscaleB}
\eea
\ie by a naive rescaling of $\alphaS^5\left(\barmudyn\right)$.

We have verified that this property is fulfilled with percent-level accuracy
by all LO curves of \reffi{fig:scanNOTaligned}.
This means that, at the level of the integrated cross section, the various scales 
\refeq{eq:murdef},\refeq{eq:murmbb}--\refeq{eq:murhbb} are
equivalent to each other.
More precisely, the scale dependence of $\sigma_\LO$ with a given dynamic scale $\mudyni{k}$
can be related to the one of a fixed scale $\mu_0$ by means of a constant
rescaling 
\bea
\mudyni{k}(\Phi)  
&\to& 
\tildemudyni{k}(\Phi) 
\,=\, \chi_k\,
\mudyni{k}(\Phi)\,,
\qquad\mbox{with}\quad
\chi_k=\frac{\mu_0}{\barmudyni{k}}\,,
\label{eq:LOalignmentA}
\eea
which results into
\bea
\sigma_\LO(\xi \tildemudyni{k}) &\simeq& \sigma_\LO(\xi \mu_0)\,.
\label{eq:LOalignmentB}
\eea
Therefore, as far as the scale uncertainty of $\sigma_\LO$ and its
normalisation are concerned, comparing different types of dynamic scales
has no significant added value wrt simple $\xir$-rescalings.
For this reason, we advocate the usage of ``aligned'' dynamic scales
$\tildemudyni{k}$, as
defined in \refeq{eq:LOalignmentA}.
In this way, the various dynamic scales have the same average value, and the
uncertainties related to this common average value are accounted for 
by standard $\xir$-rescalings, while the comparison of different scale definitions
allows one to highlight the genuine kinematic effects that are inherent in their dynamic
nature.
Comparing aligned dynamic scales yields no significant effect at 
the level of integrated cross sections, but provides 
key information on shape uncertainties, since the average
scales $\barmudyni{k}$ are sensitive both to the probed phase-space regions
and to the detailed kinematic dependence of $\mudyni{k}(\Phi)$. 
Vice versa, $\xi$-rescalings can be used to assess 
uncertainties in the normalisation of $\sigma_\LO$, whereas 
their impact on shapes is typically quite limited.

At LO, the above-mentioned alignment approach misses a crucial ingredient,
namely a good criterion for the choice of a reference scale $\mu_0$.
For $pp\to \ttbbj$, due to the very strong scale dependence induced by $\alphaS^5$, 
the choice of a well-behaved central scale is of crucial importance.
At the same time, the presence of multiple scales, distributed from 
$m_b$ to $m_{\ttbar}$ and beyond, renders this choice non-trivial.
At NLO, a natural way of addressing the scale-choice problem is to exploit
the presence of a characteristic scale given by the maximum of the NLO
scale-dependence curves, $\mumax$.
The maximum itself is not necessarily an optimal scale choice, since its
position is not guaranteed to be stable wrt higher-order corrections.  Moreover,
the flatness of the scale dependence around $\mur=\mumax$ tends to underestimate
scale uncertainties.
A more reasonable and conservative option, that will be adopted in this
paper, is to set the central scale at $\mur\simeq 2\mumax$. In this way,
the range of factor-two scale variations extends over
$[\mumax,4\,\mumax]$, covering the maximum itself as well as a relatively broad
region where $\sigma_\NLO$ is monotonically decreasing.

\begin{figure}[t!]
\centering
\includegraphics[scale=0.75]{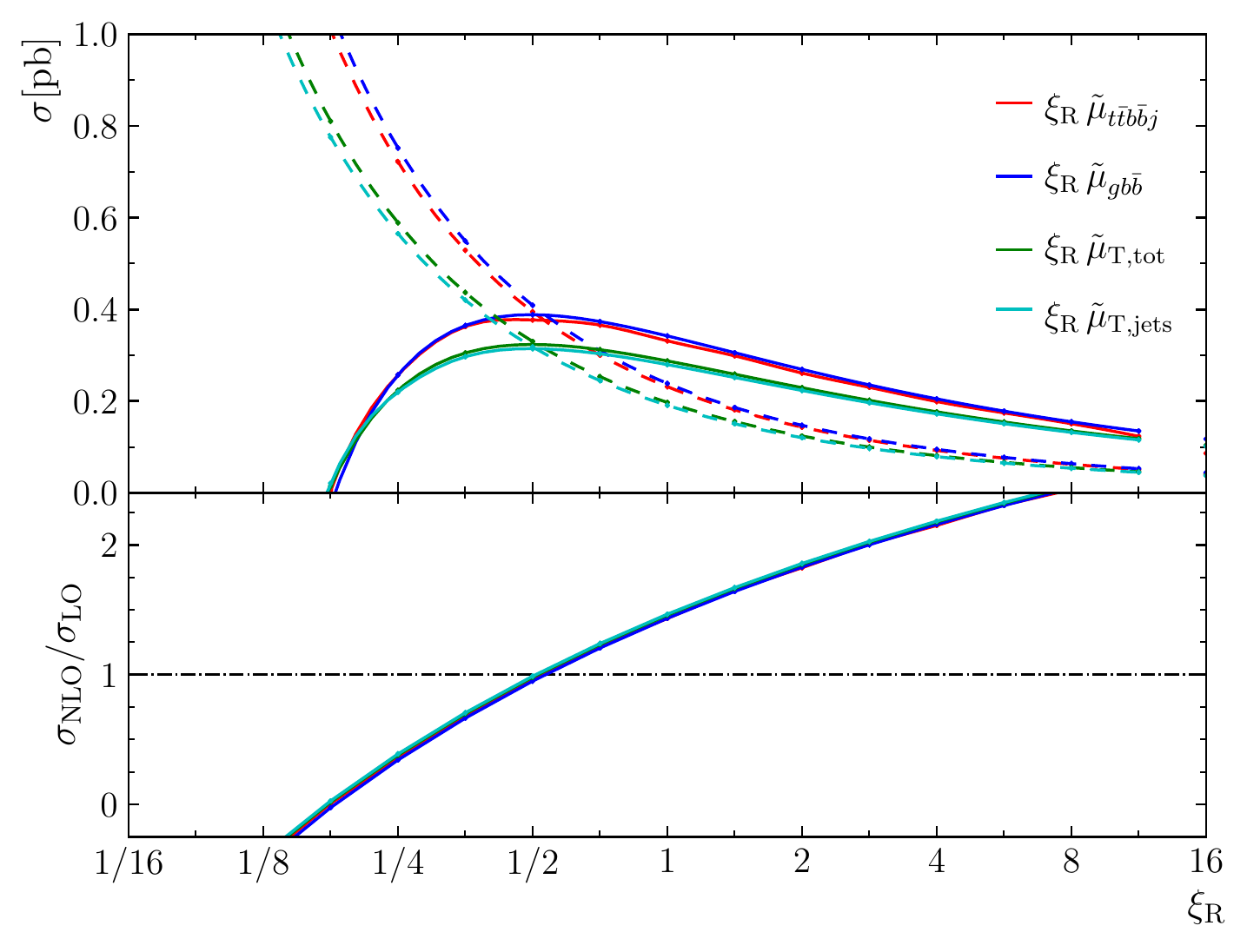}
\vspace*{-2ex}
\caption{
Renormalisation-scale dependence of the \ttbbj cross section at
$\sqrt{s}=13\,$TeV. Same predictions and variations as in 
\reffi{fig:scanNOTaligned}, but with the aligned central scales defined in~\refeq{eq:scalealignment}.
}
\vspace*{2ex}
\label{fig:scanaligned}
\end{figure}

As observed in~\reffi{fig:scanNOTaligned}, the position of $\mumax$ depends
on the choice of the dynamic scale.  However, for reasons similar to those discussed
above at LO, also NLO scale variations and the
position of their maxima can be aligned via rescalings. 
This is not entirely obvious and does not work as precisely as in the 
LO case. The main reason is that NLO cross sections consist of two kind
of contributions: Born and virtual parts, which are distributed in a similar
way as $\rd\hat\sigma_\LO$, and real-emission parts that can be distributed
in a significantly different way.
Moreover, dynamic scales can feature a different sensitivity to the 
kinematics of hard jet radiation, leading to genuinely new 
scale-dependence effects at NLO.
For these reasons, 
the LO scale-dependence model~\refeq{eq:alphadepdef}--\refeq{eq:LOalignmentB}
should be refined by splitting $\sigma_\NLO$ into two parts 
with independent average scales.
Nevertheless, for the process at hand and the scale 
choices~\refeq{eq:murdef},\refeq{eq:murmbb}--\refeq{eq:murhbb}, it turns out that 
a single overall rescaling can already yield a good level of NLO alignment. 

This is illustrated in \reffi{fig:scanaligned}, where the dynamic
scales \refeq{eq:murmbb}--\refeq{eq:murhbb}
have been rescaled in such a way that the positions of the 
NLO maxima match the maximum of $\sigma_\NLO(\mudef)$,
which is located at $0.45\,\mudef$, \ie $\mudef$ is rather close to 
$2\mumax$.
This alignment is achieved by
setting
\bea
\hatmudef &=& \mudef\,, \nonumber\\
\hatmumbb &=& 0.806\,\mumbb \,, \nonumber\\
\hatmuhtbbj &=& 1.14\,\muhtbbj\,, \nonumber\\ 
\hatmuht &=& 0.853\, \frac{\HT}{4} \,=\, 1.066\,\muht\,,
\label{eq:scalealignment}
\eea
The fact that the aligned scales are rather close to the original
choices \refeq{eq:murdef},\refeq{eq:murmbb}--\refeq{eq:murhbb} is due to the fact that the
latter had already been placed on purpose about two times above the maximum,
but without tuning their position in a precise way.
As a result of the alignment of
the NLO maxima, in \reffi{fig:scanaligned} we observe that 
the predictions based on the two  scales that depend on the jet transverse energy, \ie
$\muht$ and $\muhtbbj$, overlap almost perfectly, both at LO and NLO.
A similarly good alignment is observed also between the other two scales,
$\mudef$ and $\mumbb$, which do not depend on $\HT$.  Vice versa, the
scales that do and do not depend on $\HT$ feature a non-negligible
difference.  In particular, the values of $\sigma_\NLO $ at the maxima
differ by about 10\%.
Such differences are most likely due to the fact that the dependence on $\HT$,
which is sensitive to NLO radiation, leads to a significant difference
between the average scales in Born-like and real-emission contributions
at NLO.
Nevertheless, we observe that for all curves the position of the maximum
coincides quite precisely with the intersection of the NLO and LO curves,
which corresponds to $K=1$.  Moreover, the four $K$-factors coincide almost
exactly in the whole $\xir$ range.

In summary, applying a rescaling that aligns dynamic scales based on the 
 positions of the NLO maxima makes it possible to remove trivial differences
related to the scale normalisation and to highlight genuine differences
related to their kinematic dependence.
Since such alignment is in part already realised in the original scale
choices~\refeq{eq:murdef},\refeq{eq:murmbb}--\refeq{eq:murhbb}, in the following we will
refrain from applying the small extra rescaling~\refeq{eq:scalealignment}.

\renewcommand{\arraystretch}{1.3}
\begin{table}[t]
\vspace*{0.3ex}
\begin{center}
\begin{tabular}{c|cc|ccc|ccc}
\hline
&&&&&&\\[-3ex]
\muR & {$\Nbmin$}   & $\Njmin$  & \sigmaLO[pb]              &
$K$    &
\sigmaNLO[pb]           &
$\dfrac{\sigmaLO^{(\Njmin)}}{\sigmaLO^{(\Njmin-1)}}$                            &
$\dfrac{\sigmaNLO^{(\Njmin)}}{\sigmaNLO^{(\Njmin-1)}}$
\\[3ex]
\hline
\regionttb &1 & 0 &  $3.951^{+73\%}_{-39\%}$     & 1.92 & $7.58^{+32\%}_{-27\%}$  &     &       \\
\regionttbb &2 & 0 &  $0.3738^{+70\%}_{-38\%}$    & 1.80 & $0.674^{+27\%}_{-25\%}$ &     &       \\\hline
\regionttbj &1 & 1 &  $2.166^{+97\%}_{-45\%}$     & 1.56 & $3.38^{+21\%}_{-27\%}$  & 0.55 & 0.45 \\ 
\regionttbbj &2 & 1 &  $0.2316^{+92\%}_{-45\%}$    & 1.45 & $0.337^{+15\%}_{-25\%}$ & 0.62 & 0.50 \\\hline
\regionttbjj &1 & 2 &  $0.7812^{+119\%}_{-51\%}$   &    &     &   0.36  &     \\
\regionttbbjj &2 & 2 &  $0.08711^{+113\%}_{-50\%}$  &    &     &   0.38  &     \\\hline
\end{tabular}
\end{center}
\caption{
Cross sections at LO and NLO for $pp\to \ttbb+$jets at $\sqrt{s}=13\,$TeV.
Results are shown for integrated regions with different numbers of
$b$-jets, $\Nbmin=1,2$,
and extra light jets, $\Njmin=0,1,2$.
The acceptance cuts \refeq{eq:jetcuts} are applied.
Predictions for $\Njmin=0$ are based on a \ttbb calculation with 
$\mur=\muttbb$ and $\muf=\HT/2$, while $\Njmin=1,2$ cross sections
correspond to a \ttbbj NLO calculation with $\mur=\mudef$ and
$\muf=\HT/2$.
Seven-point scale variations are quoted in percent. In the last two
columns, for $\Njmin=1,2$ we report the ratios of LO (or NLO) cross sections
with $\Njmin$ and $(\Njmin-1)$.
The numerators and denominators of such ratios are computed at the same order.
}
\label{tab:XSA}
\end{table}

\subsection{Fiducial cross sections}
\label{se:pertplots}

In this section we present detailed numerical results for integrated cross
sections and scale uncertainties.  

To highlight the quantitative importance of light-jet radiation emitted by
the \ttbb system, in~\refta{tab:XSA} we present \ttbbjets cross sections
with variable $b$-jet and light-jet multiplicities.
Comparing the cross sections in the \regionttbbj and \regionttbb phase spaces, 
both available at NLO, we observe that the production rate for
an extra light jet is around 50\%, \ie every second \ttbb event involves a
hard light jet with $p_\rT>50$\,GeV.
The ratio of the cross sections in the \regionttbbjj and \regionttbbj regions is around
40\%, \ie the emission of a second extra jet seems to be less abundant. 
However, one should keep in mind that this ratio is only LO accurate.
The light-jet emission rates observed in the phase space 
with $\Nbmin=1$ are comparably large to the $\Nbmin=2$ case.
For fixed $\Njmin$, cross sections with two $b$-jets are 
about a factor ten smaller wrt the corresponding cross sections 
with one $b$-jet.
In general, LO scale uncertainties are very large, and grow by roughly 20\%
at each extra emission.  Instead, scale uncertainties at NLO are drastically
reduced, and in \ttbbj production they are less
pronounced than in \ttbb production.

\renewcommand{\arraystretch}{1.3}
\begin{table}[t]
  \vspace*{0.3ex}
  \begin{center}
\begin{tabular}{c|c|ccc|ll}
\hline
&&&&&&\\[-3ex]
\muR & {$\Nbmin$}   & \sigmaLO[pb]              &
$K$    &
\sigmaNLO[pb]           &
$\dfrac{\sigmaLO}{\sigmaLOdef}-1$                            &
$\dfrac{\sigmaNLO}{\sigmaNLOdef}-1$
\\[2ex]
\hline
%
{\mudef} &  1& \multicolumn{1}{c}{$2.166^{+97\%}_{-45\%}$}  & \multicolumn{1}{c}{1.56} &\multicolumn{1}{c|}{$3.38^{+21\%}_{-27\%}$} & \multicolumn{1}{c}{$0\%^{+97\%}_{-45\%}$} & \multicolumn{1}{c}{$0\%^{+21\%}_{-27\%}$} \\
& 2    & \multicolumn{1}{c}{$0.2316^{+92\%}_{-45\%}$} & \multicolumn{1}{c}{1.45} &\multicolumn{1}{c|}{$0.337^{+15\%}_{-25\%}$} & \multicolumn{1}{c}{$0\%^{+92\%}_{-45\%}$} & \multicolumn{1}{c}{$0\%^{+15\%}_{-25\%}$} \\ \hline
%
{\mumbb} & 1 &\multicolumn{1}{c}{$1.943^{+93\%}_{-45\%}$}  & \multicolumn{1}{c}{1.62} &\multicolumn{1}{c|}{$3.15^{+23\%}_{-28\%}$} & \multicolumn{1}{c}{$-10\%^{+74\%}_{-51\%}$} & \multicolumn{1}{c}{$-7\%^{+14\%}_{-33\%}$} \\
 &  2& \multicolumn{1}{c}{$0.2041^{+89\%}_{-44\%}$} & \multicolumn{1}{c}{1.56} &\multicolumn{1}{c|}{$0.318^{+19\%}_{-26\%}$} & \multicolumn{1}{c}{$-11\%^{+67\%}_{-51\%}$} & \multicolumn{1}{c}{$-6\%^{+12\%}_{-30\%}$} \\ \hline
%
{\muhtbbj} &1    & \multicolumn{1}{c}{$1.772^{+91\%}_{-44\%}$}  & \multicolumn{1}{c}{1.51} &\multicolumn{1}{c|}{$2.68^{+15\%}_{-25\%}$} & \multicolumn{1}{c}{$-18\%^{+56\%}_{-54\%}$} & \multicolumn{1}{c}{$-21\%^{-9\%}_{-41\%}$} \\
\multicolumn{1}{c|}{} &2       & \multicolumn{1}{c}{$0.2100^{+90\%}_{-44\%}$} & \multicolumn{1}{c}{1.37} &\multicolumn{1}{c|}{$0.287^{+7\%}_{-22\%}$} & \multicolumn{1}{c}{$-9\%^{+72\%}_{-49\%}$} & \multicolumn{1}{c}{$-15\%^{-8\%}_{-34\%}$} \\ \hline
%
{\muht}  &1  & \multicolumn{1}{c}{$1.697^{+90\%}_{-44\%}$}  & \multicolumn{1}{c}{1.60} &\multicolumn{1}{c|}{$2.71^{+19\%}_{-26\%}$} & \multicolumn{1}{c}{$-22\%^{+49\%}_{-56\%}$} & \multicolumn{1}{c}{$-20\%^{-4\%}_{-41\%}$} \\
  & 2&  \multicolumn{1}{c}{$0.2064^{+89\%}_{-44\%}$} & \multicolumn{1}{c}{1.41} &\multicolumn{1}{c|}{$0.291^{+10\%}_{-23\%}$} & \multicolumn{1}{c}{$-11\%^{+69\%}_{-50\%}$} & \multicolumn{1}{c}{$-13\%^{-5\%}_{-34\%}$} \\ \hline
\end{tabular}
\end{center}
\caption{Cross sections at LO and NLO for $pp\to \ttbbj$ at $\sqrt{s}=13\,$TeV.
Results are shown for the fiducial regions with $\Njmin=1$ and
$\Nbmin=1$ (\regionttbj) or $\Nbmin=2$ (\regionttbbj). The acceptance cuts~\refeq{eq:jetcuts}
are applied.
Four different choices of $\mur$ as defined in
\refeq{eq:murdef},\refeq{eq:murmbb}--\refeq{eq:murhbb} are compared, while the factorisation scale
\refeq{eq:muf} is used throughout.
Columns 3--5 show absolute predictions at LO and NLO, 
as well as the usual correction factor $K=\sigmaNLO/\sigmaLO$.
Uncertainties given in percent correspond to 
seven-point factor-two variations of $\mur$ and 
$\muf$. %
Columns 6 and 7 show the relative differences between LO and NLO
cross sections, respectively, based on the default  $\mur=\mudef$ 
and the other dynamical scales.
As central values we report ratios obtained with the nominal
values of the various scales, while lower\,(upper) values correspond to the
minimum\,(maximum) of
$\sigma_{\mathrm{(N)LO}}(\xir,\xif)/\sigma_{\mathrm{(N)LO, def}}-1$,
where seven-point variations are restricted to the numerator.
The reported cross sections have been computed with Monte Carlo statistical
uncertainties at the level of three permille at NLO and below one
permille at LO.
}
\vspace*{1ex}
\label{tab:XS}
\end{table}

In the following we focus on LO and NLO predictions for $\ttbb+$jet
production in the \regionttbj and \regionttbbj phase-space regions.
In \refta{tab:XS} we compare cross sections and scale variations based on
the four dynamic scale choices~\refeq{eq:murdef},\refeq{eq:murmbb}--\refeq{eq:murhbb}.
For what concerns nominal predictions (without scale variations), the
default scale $\mur=\mudef$ yields the largest cross sections.  
At LO, the other predictions are between 10\% and 20\% lower. The 
\regionttbbj\!(\regionttbj) cross sections based on the $\HT$-dependent scales remain
15\% (20\%) lower also at NLO.
In contrast, the two $\HT$-independent scales agree at the level of 5\% at
NLO.
Comparing the cross sections with one and two $b$-jets, using
$\HT$-independent scales we observe a ratio very close to $1/10$, while the
other scale choices yield a ratio of $1/9.3$.

Seven-point scale variations at LO are between around $-45\%$ and $+90\%$ for all
scale choices, both in the \regionttbj and \regionttbbj regions.
At NLO they are reduced around 20\%, with significant differences depending
on the scale choice and the number of $b$-jets.
In the \regionttbbj\!(\regionttbj) phase space, the half-width of the scale-variation band
is around 20\%\,(25\%) for the $\HT$-independent scales and about 5\%
smaller for the $\HT$-dependent ones.

In the last two columns of \refta{tab:XS}, we compare LO and NLO cross
sections and seven-point variations of the various dynamic scales,
normalising the results to nominal predictions with the default scale choice.
The scale-variations bands obtained with the $\HT$-dependent scales are
significantly lower than the other bands.
At NLO, the variation of the default scale covers the absolute NLO maximum
observed in \reffi{fig:scanNOTaligned}, while the upper variations of the
$\HT$-dependent scales are \percentrange{20}{30} lower.
Vice versa, the lower variation of the default scale is \percentrange{10}{15} above the
corresponding variation of the $\HT$-dependent scales.
In the \regionttbbj\!(\regionttbj) phase spaces, the variations of the 
default scale change the nominal cross section by
$[-25\,(27)\%,+15\,(21)\%]$, while the envelope of the four variation bands
corresponds to $[-34\,(41)\%,+15\,(21)\%]$, which amounts to an increase
of the half-band width from 20\,(24)\% to 25\,(31)\%, \ie by 5\,(7)\%.
Vice versa, using the $\muht$ result as a reference gives a $\muht$
variation band of $[-23\,(26)\%,+10\,(19)\%]$ and an envelope band of
$[-23\,(26)\%,+32\,(51)\%]$, which corresponds to
an increase of the half-band width from 
17\,(23)\% to 28\,(38\%)\%, \ie by 11\,(15)\%.
We also note that the variation bands of the $\HT$-independent scales cover
the nominal predictions of the $\HT$-dependent scales, but not vice versa.
Based on this observations, we conclude that the somewhat larger seven-point
variation of the $\HT$-independent scales should be regarded as a more
realistic estimate of scale uncertainties.

\renewcommand{\arraystretch}{1.3}
\begin{table}[t]
\vspace*{0.3ex}
\begin{center}
\begin{tabular}{c|c|ccc|ll}
\hline
&&&&&&\\[-3ex]
\muR & {$\Nbmin$}   & \sigmaLO[pb]              &
$K$    &
\sigmaNLO[pb]           &
$\dfrac{\sigmaLO}{\sigmaLOdef}-1$                            &
$\dfrac{\sigmaNLO}{\sigmaNLOdef}-1$
\\[2ex]
\hline
%
{\hatmudef} & 1& \multicolumn{1}{c}{$2.166^{+97\%}_{-45\%}$}  & \multicolumn{1}{c}{1.56} &\multicolumn{1}{c|}{$3.38^{+21\%}_{-27\%}$} & \multicolumn{1}{c}{$0\%^{+97\%}_{-45\%}$} & \multicolumn{1}{c}{$0\%^{+21\%}_{-27\%}$} \\
& 2& \multicolumn{1}{c}{$0.2316^{+92\%}_{-45\%}$} & \multicolumn{1}{c}{1.45} &\multicolumn{1}{c|}{$0.337^{+15\%}_{-25\%}$}  & \multicolumn{1}{c}{$0\%^{+92\%}_{-45\%}$} & \multicolumn{1}{c}{$0\%^{+15\%}_{-25\%}$} \\ \hline
%
{\hatmumbb } & 1&  \multicolumn{1}{c}{$2.291^{+97\%}_{-46\%}$}  & \multicolumn{1}{c}{1.48} &\multicolumn{1}{c|}{$3.40^{+16\%}_{-26\%}$} & \multicolumn{1}{c}{$+6\%^{+109\%}_{-43\%}$} & \multicolumn{1}{c}{$+0.4\%^{+17\%}_{-26\%}$} \\
& 2& \multicolumn{1}{c}{$0.2388^{+93\%}_{-45\%}$} & \multicolumn{1}{c}{1.42} &\multicolumn{1}{c|}{$0.338^{+13\%}_{-24\%}$}  & \multicolumn{1}{c}{$+3\%^{+99\%}_{-43\%}$} & \multicolumn{1}{c}{$+0.3\%^{+13\%}_{-24\%}$} \\ \hline
%
{\hatmuhtbbj}   & 1&  \multicolumn{1}{c}{$1.606^{+89\%}_{-44\%}$}  & \multicolumn{1}{c}{1.60} &\multicolumn{1}{c|}{$2.57^{+19\%}_{-26\%}$}   & \multicolumn{1}{c}{$-26\%^{+40\%}_{-58\%}$} & \multicolumn{1}{c}{$-24\%^{-9\%}_{-43\%}$} \\
& 2& \multicolumn{1}{c}{$0.1909^{+88\%}_{-44\%}$} & \multicolumn{1}{c}{1.45} &\multicolumn{1}{c|}{$0.277^{+12\%}_{-24\%}$}   & \multicolumn{1}{c}{$-18\%^{+54\%}_{-53\%}$} & \multicolumn{1}{c}{$-18\%^{-8\%}_{-37\%}$} \\ \hline
%
{\hatmuht}  & 1&  \multicolumn{1}{c}{$1.621^{+89\%}_{-44\%}$}  & \multicolumn{1}{c}{1.64} &\multicolumn{1}{c|}{$2.65^{+21\%}_{-27\%}$} & \multicolumn{1}{c}{$-25\%^{+41\%}_{-58\%}$} & \multicolumn{1}{c}{$-21\%^{-5\%}_{-43\%}$} \\
& 2& \multicolumn{1}{c}{$0.1973^{+88\%}_{-44\%}$} & \multicolumn{1}{c}{1.44} &\multicolumn{1}{c|}{$0.285^{+12\%}_{-24\%}$}   & \multicolumn{1}{c}{$-15\%^{+60\%}_{-52\%}$} & \multicolumn{1}{c}{$-15\%^{-5\%}_{-36\%}$} \\ \hline
\end{tabular}
\end{center}
\caption{
Cross sections at LO and NLO for $pp\to \ttbbj$ at $\sqrt{s}=13\,$TeV in the
\regionttbj ($\Nbmin=1$) and \regionttbbj ($\Nbmin=2$) phase spaces.
Similar predictions and variations as in \refta{tab:XS} for the case of the
aligned central scales defined in \refeq{eq:scalealignment}.
}
\vspace*{1ex}
\label{tab:alignedXS}
\end{table}

In \refta{tab:alignedXS} we present similar results based on the aligned
scales \refeq{eq:scalealignment}, which correspond to
\reffi{fig:scanaligned}.
The main effect of the alignment is that the LO and NLO cross sections based
on the two $\HT$-independent scales become much closer to
each other, while predictions based on the $\HT$-dependent scales change in a
less significant way.
This is mainly due to the fact the original scales $\muhtbbj$ and $\muht$
are already very close to the corresponding aligned scales
in~\refeq{eq:scalealignment}.
In any case, predictions based on the aligned scales are
independent of the initial normalisation of the various scales.

After alignment, we still see significant differences between
the predictions with $\HT$-dependent and $\HT$-independent scales.
More precisely, due to the fact that the alignment is based on the NLO
maximum of the \regionttbbj cross sections, the spread between $K$-factors in the
\regionttbbj phase space  goes down from $0.11$ to $0.03$.
Vice versa, the $K$-factor difference in the \regionttbj phase space increases from
$0.11$ to $0.16$.
The alignment leads also to a slight reduction of NLO scale uncertainties,
and the nominal predictions based on $\HT$-independent scales remain above
the NLO bands of $\HT$-dependent scales.
Such differences between aligned NLO predictions in different phase-space
regions should be regarded as genuine effects of the kinematic dependence of
dynamic scales.  Thus they play a largely complementary role wrt factor-two
scale variations.

\subsection{Sudakov effects}
\label{se:sudakov}

\begin{figure}[t]
\bce
\includegraphics[scale=0.56,trim=0 0 0 0,clip]{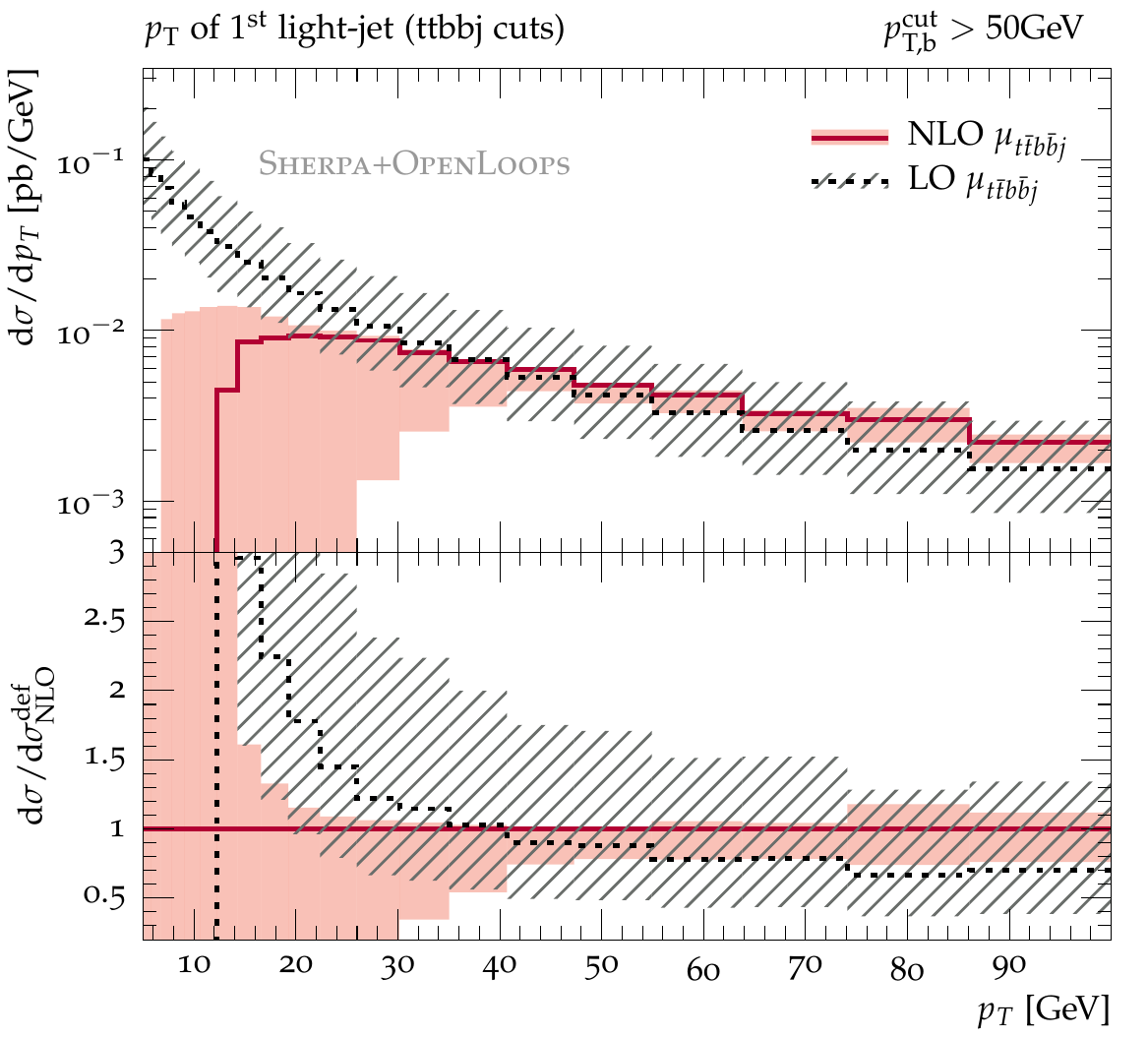}
\hspace{6mm}
\includegraphics[scale=0.56,trim=0 0 0 0,clip]{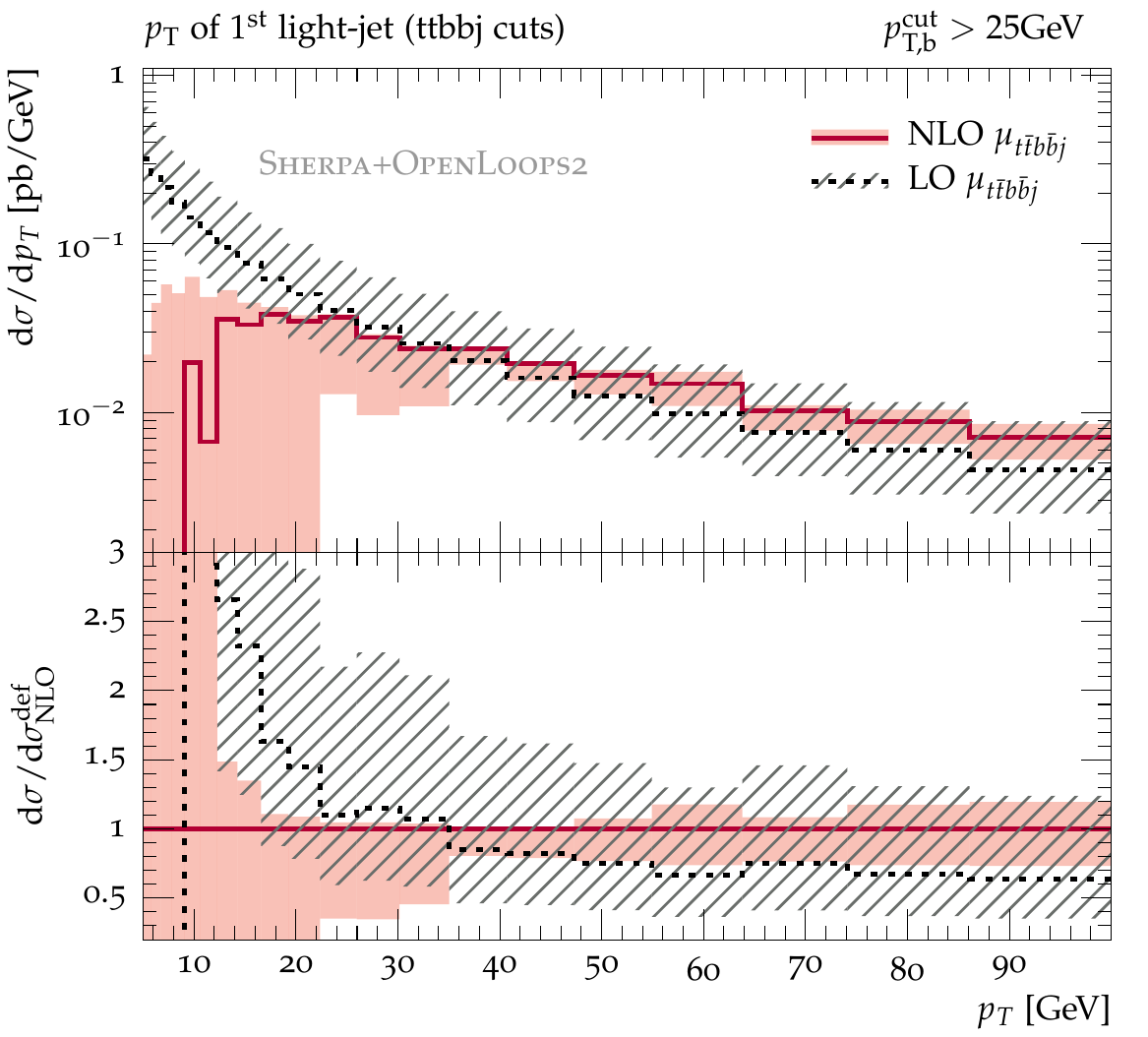}
\ece
\label{fig:sud25:2_logPT_J1_lowpt}
\vspace*{-3.5ex}
\caption{%
Distribution in the $\pT$ of the leading light jet in the soft region for
$pp\to \ttbbj$ in the \regionttbbj phase space.
In the left (right) plot $b$-jets are subject to the usual cuts with
standard (reduced) $\pTb$ threshold of $50$\,GeV\,(25\,GeV).  Cuts on the
jet-$\pT$ have been lowered to 5\,\GeV, and all jets are subject to a
pseudo-rapidity cut $|\eta|<2.5$.
The upper frames show LO (dashed) and NLO (solid) absolute predictions with
the default scale choice, $\mur=\mudef$ and $\muf=\HT/2$.  The bands
correspond to seven-point scale variations.
The ratio plots display relative differences wrt to the nominal NLO
predictions.
}
\label{fig:sud:2_logPT_J1_lowpt}
\vspace*{2ex}
\end{figure}

In this section we address the question of the safeness of the 
chosen transverse-momentum cut of 50\,GeV 
with respect to higher-order Sudakov logarithms.
To investigate such Sudakov effects, which 
appear in the region where the $\pT$ of the light jet, $\pTj$,
becomes small,
we relax the cut on $\pTj$ and, in~\reffi{fig:sud:2_logPT_J1_lowpt},
we study the perturbative behaviour of
the $\rd\sigma/\rd\pTj$ distribution.
In the left plot,  this is done by keeping the usual $b$-jet cuts
at $\pTcut=50\,$GeV, while in the right plot this threshold
is lowered to $\pTcut=25\,$GeV.

As is well known, the $\rd\sigma/\rd\pTj$ distribution is logarithmically
divergent at LO, while summing such logarithms to all orders in $\alphaS$ 
would cancel the divergence and lead to $\rd\sigma/\rd\pT\to 0$ at
small $\pT$.
In the fixed-order NLO calculation at hand, this behaviour manifests itself through
an increasingly strong shape difference between the LO and NLO distributions
at small $\pT$.  
For $\pTcut=50\,$GeV, we find that at $\pT\simeq
20\,$GeV the NLO curve develops a Sudakov peak, below which NLO corrections
start overcompensating the logarithmic growth of the LO distribution.
In correspondence with the Sudakov peak, the NLO cross section is
already less than half of the LO one, and below 15\,GeV it rapidly falls
into the unphysical regime of negative cross sections.
This pathologic behaviour of the fixed-order NLO prediction is also
reflected by the rapid inflation of NLO scale uncertainties below 40\,GeV,
while our choice of setting the light-jet $\pT$ cuts at 50\,GeV guarantees
good stability both for the NLO predictions and their uncertainties.

\renewcommand{\arraystretch}{1.3}
\begin{table}[t]
\vspace*{0.3ex}
\begin{center}
\begin{tabular}{c|ccc|ccc}
\hline
$\pTcut$[GeV] &  & 50 &  &  & 25 &  \\\hline
$p_{\rT,j}$[GeV] & 25 & 50 & 100  & 25 & 50 & 100 \\\hline
&&&&&&\\[-3ex]
$\dfrac{\rd\sigma_\LO/\rd\pTj}{\rd\sigma_\NLO/\rd\pTj}$ & 1.45 & 0.881 & 0.699 & 1.09 & 0.754 & 0.639 \\[2ex]
\hline
\end{tabular}
\end{center}
\caption{%
Comparison of the LO and NLO distributions in the leading-jet $\pT$ for 
$\pTcut=50$\,GeV and $25$\,GeV. 
The results correspond to the \regionttbbj phase space 
with the cut $\pT>\pTcut$ restricted to $b$-jets.
The reported values at $\pTj/\mathrm{GeV}=25, 50, 100$ correspond to the
bins $\left[22.4,26.0\right]$, $\left[47.3,55.0\right]$ and
$\left[86.0,100\right]$.
}
\label{tab:sudakovratios}
\end{table}

As can be seen in the right plot of \reffi{fig:sud:2_logPT_J1_lowpt},
reducing the $b$-jet threshold to 25\,GeV tends to lower the position of the
Sudakov peak by 5\,GeV or so.
In this case, NLO predictions feature a good perturbative convergence down to
30--35 \,GeV.
The effect of NLO corrections on the jet-$\pT$ distribution 
for selected values of $\pT$ is reported in~\refta{tab:sudakovratios}.

\section{Differential observables}
\label{se:ttbbjdist}

In this section we study differential observables for $pp\to \ttbbj$ at
13\,TeV restricting ourselves to the \regionttbbj phase space.  The main focus of
our analysis is on the shapes of distributions and related uncertainties.

\subsection{Distributions and shape uncertainties in the \regionttbbj phase space}
\label{se:distributions}

In \reffis{fig:2_PT_J1}{fig:2_M_B1B2} we analyse a series of differential
distributions showing, for each observable, absolute and normalised
distributions as well as six different ratio plots, which quantify the
relative effects of seven-point variations and differences between the
various dynamic scales.
We restrict ourselves to the three dynamic
scales~\refeq{eq:murdef},\refeq{eq:murmbb}--\refeq{eq:murht}, since including or not the 
scale~\refeq{eq:murhbb} does not change the overall picture of shape
uncertainties.
The format of the plots is described in the following and in the caption of \reffi{fig:2_PT_J1},
and it is the same for all figures in this section.

The left plot of each Figure contains:
\begin{itemize}
\item[(L1)] An upper frame with LO and NLO distributions based on the
default scale choice $(\mur,\muF)=(\mudef, \HT/2)$, as well as the
corresponding seven-point variation bands.
\item[(L2)] A first ratio plot corresponding to the inverse $K$-factor,
\bea
K^{-1}_{\mathrm{(N)LO}}(\xir,\xif) &=&
\frac{\sigma_{\mathrm{(N)LO}}(\xir\,\mudef,\xif\,\muF)}
{\sigma_{\mathrm{NLO}}(\mudef,\muf)}\,,
\label{eq:invabskfactor}
\eea
where scale variations are applied only in the numerator.
\item[(L3)] A second ratio plot that features the LO and NLO ratios,
\bea
R_{\mathrm{(N)LO}}(\mur)&=&
\frac{\sigma_{\mathrm{(N)LO}}(\xir\,\mur,\xif\,\muF)}
{\sigma_{\mathrm{(N)LO}}(\xir\,\mudef,\xif\,\muF)}\,.
\label{eq:absshaperatio}
\eea
This ratio encodes differences between the dynamic scale $\mur=\mumbb$, 
defined in~\refeq{eq:murmbb}, and the default scale.
Seven-point scale variations are applied in a correlated way 
to the numerator and the denominator.
In this way, the main effect of factor-two variations, which amounts to 
a nearly constant normalisation shift,
%
%
cancels out.
As a result, the ratio \refeq{eq:absshaperatio} is mostly sensitive to effects that 
arise from the different kinematic dependence of the considered scales, and cannot
be accounted for by factor-two variations of a single scale.
\item[(L4)] A third ratio plot that shows the ratio
\refeq{eq:absshaperatio} for $\mur=\muht$.
\eit

\noindent The right plot of each figure shows the following normalised
distributions and ratios thereof.
\bit
\item[(R1)] The upper frame displays the LO and NLO normalised
distributions,
\bea
\rd\hat\sigma_{\mathrm{(N)LO}}(\xir\,\mur,\xif\,\muf)&=&
\frac{\rd\sigma_{\mathrm{(N)LO}}(\xir\,\mur,\xif\,\muf)}{\sigma_{\mathrm{(N)LO}}(\xir\,\mur,\xir\,\muf)}\,,
\label{eq:normdist}
\eea
for the default scale $\mur=\mudef$.
The denominator corresponds to the integrated cross section in the \regionttbbj phase
space, and seven-point variations in the numerator and denominator are
correlated.  
In this way, distributions are always normalised to one, \ie normalisation
effects cancel out, and only shape corrections and uncertainties remain visible.
\item[(R2)] 
The first ratio plot shows the ratio of normalised distributions,
\bea
\hat R_{\mathrm{(N)LO}}(\mudef)&=&
\frac{\rd\hat\sigma_{\mathrm{(N)LO}}(\xir\,\mudef,\xif\,\muf)}
{\rd\hat\sigma_{\mathrm{NLO}}(\mudef,\muf)}\,,
\label{eq:normshapedist}
\eea
based on the default scale.
Here seven-point variations are applied only to the numerator, but 
their normalisation effect cancels out as in~\refeq{eq:normdist}.
Thus the ratio \refeq{eq:normshapedist} highlights the relative effect of
NLO corrections and seven-point variations on the shape of distributions.
\item[(R3)]
The second ratio plot shows the ratios of normalised distributions at LO, 
\bea
\overline R_{\mathrm{LO}}(\mur)&=&
\frac{\rd\hat\sigma_{\mathrm{LO}}(\xir\,\mur,\xif\,\muf)}
{\rd\hat\sigma_{\mathrm{LO}}(\mudef,\muf)}\,,
\label{eq:normshapedistBLO}
\eea
for the three dynamic scales $\mur=\mudef$, $\mumbb$, $\muht$. 
This ratio highlights shape differences between those scales (with
seven-point variations) and the nominal default scale.
\item[(R4)] The third ratio plot shows the same ratios as defined in~\refeq{eq:normshapedistBLO}, but at NLO, 
\bea
\overline R_{\mathrm{NLO}}(\mur)&=&
\frac{\rd\hat\sigma_{\mathrm{NLO}}(\xir\,\mur,\xif\,\muf)}
{\rd\hat\sigma_{\mathrm{NLO}}(\mudef,\muf)}\,,
\label{eq:normshapedistBNLO}
\eea
for $\mur=\mudef$, $\mumbb$, $\muht$. 
\eit


\begin{figure}[t]
\showscanplots{2_PT_J1}
\caption{Distribution in the $\pT$ of the leading light jet
for $pp\to \ttbbj$ at 13\,TeV in the \regionttbbj phase space with acceptance cuts
\refeq{eq:jetcuts}. 
The left figure shows LO (dashed) and NLO (solid) absolute predictions and
ratios thereof. The bands correspond to seven-point scale variations.
The upper frame (L1) displays the absolute $\pT$ distribution with
$\mur=\mudef$, and the first ratio plot (L2) shows the corresponding
(inverse) $K$-factor defined in~\refeq{eq:invabskfactor}.
The other ratio plots on the left display the ratios $R_{\mathrm{(N)LO}}(\mur)$, defined
in~\refeq{eq:absshaperatio} for the scales $\mur=\mumbb$ (L3) and $\muht$
(L4).
Such ratios quantify shape uncertainties due to the differences between the
default scale and the alternative dynamic scales.  Seven-point variations in
the numerator and the denominator are correlated.
The right plots present normalised distributions and ratios thereof.
The upper frame (R1) shows the LO and NLO normalised distributions
\refeq{eq:normdist} based on the default scale, with correlated seven-point
variations in the numerator and denominator.
The first ratio plot (R2) displays the ratio $\hat
R_{\mathrm{(N)LO}}(\mudef)$, which is defined in~\refeq{eq:normshapedist}
and highlights the relative shape distortions induced by NLO corrections and
scale variations.
The last two ratio plots on the right feature the ratios $\overline
R_{\mathrm{(N)LO}}(\mur)$ for $\mur=\mudef$, $\mumbb$ and $\muht$ at LO (R3)
and NLO (R4).
As defined in~\refeq{eq:normshapedistBLO}--\refeq{eq:normshapedistBNLO},
such ratios quantify shape uncertainties associated with the kinematic
dependence of the different dynamic scales.
}
\vspace*{3ex}
\label{fig:2_PT_J1}
\end{figure}
\def\samesetup{Same setup and plots as in \reffi{fig:2_PT_J1}.}

\reffi{fig:2_PT_J1} presents the distribution in the $\pT$ of the leading light
jet up to 400\,GeV.
The corrections to the shape of this distribution indicate excellent
perturbative stability in the hard region above 150\,GeV: The default scale
yields a nearly constant $K$-factor around $1.65$, and the 
scale-variation band is also quite stable at the $\pm 20\%$ level.
In the region below 150\,GeV, as already observed
in~\reffi{fig:sud:2_logPT_J1_lowpt}, NLO effects start affecting the
$\pT$-shape with a correction of about 25\% between 150 and 50\,GeV.
Such effects can be attributed to Sudakov logarithms, and estimating the
missing higher-order corrections via naive exponentiation, we expect residual
shape uncertainties below 5\% at NLO.

Comparing predictions based on the default scale 
and the other dynamic scales, in L3--L4 we observe normalisation
differences at the level of \percentrange{10}{15}, which are compatible with the NLO
scale-variation band in L2. 
These differences are very stable with respect
to correlated factor-two scale variations as defined in 
\refeq{eq:normshapedistBLO}: at LO such variations cancel 
almost exactly, and also the NLO bands in L3--L4 are suppressed at the 
level of 5\% or less.
Comparing normalised distributions with different dynamic scales in 
R3--R4, we see that LO shapes (and their seven-point variations) 
are almost identical, with only few-percent 
differences between $\muht$ and the $\HT$-independent scales.
The nominal NLO predictions based on the various scales feature a
similarly high level of agreement (see R4).
However, similarly as in R2, factor-two
variations lead to shape distortions at the 20\% level.
Such distortions shift the shape of the distributions in the region
below 150\,GeV, and are compensated by an opposite, but $\pT$-independent 
shift in the hard region.
In general, the suppression of shape effects at LO 
demonstrates the importance of NLO predictions 
for a more realistic assessment of shape uncertainties.


The non-negligible NLO shape effects observed in~\reffi{fig:2_PT_J1} are a
specific feature of the jet-$\pT$ distribution in the vicinity of the cut,
while other distributions that involve the leading light jet are typically
more stable.

\begin{figure}[t] 
\scanplots{2_ETA_J1}
\caption{Pseudo-rapidity of the leading light jet.  \samesetup}
\vspace*{3ex}
\label{fig:2_ETA_J1}
\end{figure}

\begin{figure}[t] 
\scanplots{2_DR_J1B1B2}
\caption{$\Delta R$ between the light jet and the $b$-jet pair.  \samesetup}
\vspace*{3ex}
\label{fig:2_DR_J1B1B2}
\end{figure}

This is illustrated in \reffis{fig:2_ETA_J1}{fig:2_DR_J1B1B2} where we
present the distributions in the pseudo-rapidity of the
leading jet and in its $\Delta R$ separation with respect to the leading 
$b$-jet.
For these observables, NLO corrections and uncertainties 
correspond to the ones of the integrated cross section 
and depend only very weakly on the jet kinematics.
In fact, as can be seen from the 
ratio plots R2--R4, the shape of such distributions turns out to be stable
at the percent level with respect of 
seven-point variations and differences between dynamic scales.

In general, as found in~\reffis{fig:2_PT_J1}{fig:2_DR_J1B1B2} and in
various other observables not shown here, distributions in the leading light
jet can be controlled with typical normalisation uncertainties of order 
20\%  and shape uncertainties of order 10\% or below.

\begin{figure}[t]
\scanplots{2_PT_T1}
\caption{$\pT$ of the harder top. \samesetup}
\vspace*{4ex}
\label{fig:2_PT_T1}
\vspace*{3ex}
\scanplots{2_PT_T2}
\caption{$\pT$ of the softer top. \samesetup}
\label{fig:2_PT_T2}
\end{figure}

\begin{figure}[t]
\scanplots{2_PT_B1}
\caption{$\pT$ of the first $b$-jet. \samesetup}
\vspace*{4ex}
\label{fig:2_PT_B1}
\vspace*{3ex}
\scanplots{2_PT_B2}
\caption{$\pT$ of the second $b$-jet. \samesetup}
\label{fig:2_PT_B2}
\end{figure}

\begin{figure}[t]
\scanplots{2_DR_B1B2}
\caption{$\Delta R$ between the two $b$-jets. \samesetup}
\vspace*{4ex}
\label{fig:2_DR_B1B2}
\vspace*{3ex}
\scanplots{2_M_B1B2}
\caption{Invariant mass of the $b$-jet pair. \samesetup}
\label{fig:2_M_B1B2}
\end{figure}


In \reffis{fig:2_PT_T1}{fig:2_M_B1B2} we present distributions in the
top-quark and $b$-jet kinematics.
For the transverse momentum of the harder top quark, shown in
\reffi{fig:2_PT_T1}, we find that NLO corrections and scale variations are
very stable, the only exception being a NLO shape correction of about $15\%$
in the region below 50\,GeV, where the cross section is strongly suppressed.
For the $\pT$ of the softer top quark,
shown in \reffi{fig:2_PT_T2}, NLO corrections feature a 
moderate, but more significant kinematic dependence.
In particular, the $K$-factor goes down from about 1.5 in the bulk of the
distribution to 1.2 in the tail, while seven-point scale variations lead to
a similarly large shape distortion in the tail (see R2, R4).
This behaviour is qualitatively quite similar to the Sudakov effects 
observed in the
soft region of the jet-$\pT$ distribution in \reffi{fig:2_PT_J1}.
It can be attributed to the
fact that requiring two very hard top quarks restricts the available phase
for additional radiation, confining the light jet 
into the soft region close to the 50\,GeV threshold.


The distributions in the $\pT$ of the harder and softer $b$-jets,
shown in \reffis{fig:2_PT_B1}{fig:2_PT_B2}, feature a qualitatively very similar
behaviour as the corresponding top-quark distributions.
In the case of the harder $b$-jet $\pT$, NLO corrections and scale
uncertainties depend rather weakly on $\pT$ (although more significantly
than for the harder top quark), while the distribution in the 
$\pT$ of the softer $b$-jet features strong NLO effects, 
which are most likely due to Sudakov logarithms.


Finally, in \reffis{fig:2_DR_B1B2}{fig:2_M_B1B2} we show the $\Delta R$
separation and the invariant-mass distribution of the $b$-jet pair.
For these observables, as far as the default scale and the scale
$\mur=\muht$ are concerned, NLO corrections and variations feature
very little kinematic dependence, with percent-level shape differences.
On the contrary, the dynamic scale $\mur=\mumbb$ leads to a very different
shape in the tail of the $\DRbb$ distribution,
with deviations that reach $-45\%$ at LO
and remain as important as $-30\%$ at NLO. 
A similar, although less dramatic trend is observed also in
the tail of the invariant-mass distribution, which is clearly correlated 
to the tail of the $\DRbb$ distribution. 
These effects are most pronounced at 
$\DRbb > \pi$, where the two $b$-jets are emitted in opposite
hemispheres. 
In this region, 
the main mechanism of \ttbb production via 
final-state $g\to\bbbar$ splittings (see \reffi{fig:ttbbtopologies}a)
is strongly suppressed, and the 
leading role is played by topologies with initial-state
$g\to \bbbar$ splittings (see \reffi{fig:ttbbtopologies}b in this paper and Fig.6 in~\cite{Jezo:2018yaf}).
The latter are maximally enhanced at 
$E_{\rT,b} \ll m_{bb}$, and their characteristic virtualities 
of order $E_{\rT,b}$ are correctly reflected
in the definition of the scales $\mudef$ and $\muht$.
Instead, the term $m_{\bbbar}$ in~\refeq{eq:murmbb}
renders $\mumbb$ unnaturally hard, leading to  
an unphysical suppression of the tails.
It is clear that this behaviour cannot be 
regarded as a theoretical uncertainty, but should simply be taken as 
an indication that
the scale $\mumbb$, which was designed to account for final-state $g\to \bbbar$
splittings, is not applicable to initial-state $g\to \bbbar$ splittings.
On the contrary,
the scales $\mudef$ and $\muht$ turn out to be well behaved for both kinds of
splittings.


\subsection{Recoil observables}
\label{se:recoil}

As pointed out in the introduction, 
the accuracy of NLO Monte Carlo simulations of \ttbb
production plays a key role in $\ttbar H$ analyses.
In this context, it was recently observed that
the modelling of recoil effect by the parton shower may be 
a dominant source of uncertainty
(see \eg \cite{talkPozzoriniHXSWGDec18,talkPozzoriniHXSWGJun19}).
This is not surprising, given that every second \ttbb event is accompanied by QCD radiation with
$\pT>50\,$GeV (see~\refta{tab:XSA}).
In fact, away from the collinear regions, the recoil prescriptions used by parton showers 
can easily lead to unphysical momentum shifts of the order of
10\,GeV and beyond.
In the case of $b$-jets the effects of recoil mismodelling can
be quite significant.
In particular, shifts in the transverse momentum of the second $b$-jet
can easily result 
in sizeable migration effects from the strongly populated
region with $\Nbmin=1$ to the less populated $\Nbmin=2$ region.\footnote{
We have verified that in the \regionttbj region 
the second $b$-jet is typically slightly below the 
$\pT$ acceptance cut and is almost ten times softer 
with respect to the leading light jet. Thus, 
a small fraction of the QCD recoil is sufficient in order
to shift the softer $b$-jet 
above the acceptance cut.
More precisely, in the \regionttbj\!(\regionttbbj)
phase space with standard cuts at 50\,GeV the average
transverse momenta of light jets and $b$-jets are
$\ptav{j_1}=131\,(137)$\,GeV, $\ptav{b_1}=134\,(166)$\,GeV and
$\ptav{b_2}=35\,(86)$\,GeV, while their average ratios are
$\ptratav{b_1}=1.34\,(1.09)$ and $\ptratav{b_2}=9.15\,(1.83)$\,.
In the \regionttbj phase space, the quoted $\ptav{b_2}$ and
$\ptratav{b_2}$ averages 
have been evaluated including only events that involve a second resolved $b$-jet 
with $p_{\rT,b_2}>0$ and $|\eta_{b_2}|<2.5$.
%
}
\begin{figure}[t]
\centering
\includegraphics[width=0.4\textwidth]{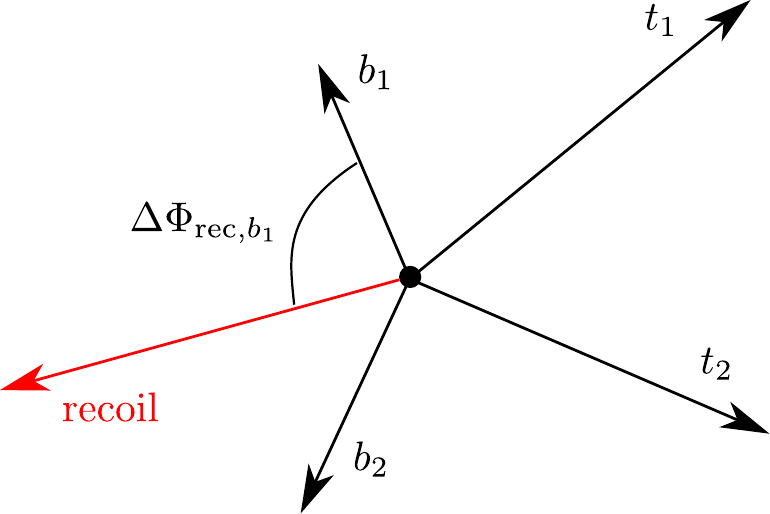}
\caption{
Sketch of the azimuthal angular correlation $\Dphirec{X}$ between individual
objects of the \ttbb system and its recoil.
See~\refeq{eq:phirecdef}--\refeq{eq:recdef}.
}
\label{fig:recoilsketch}
\end{figure}
In this context, the accurate description of QCD radiation provided by the
calculation of $pp\to \ttbbj$ at NLO can be exploited as a benchmark to test
the modelling of recoil effects in \ttbb Monte Carlo simulations.
With this motivation in mind, we study the azimuthal
angular correlations~\cite{talkPozzoriniHXSWGJun19}
\bea
\Dphirec{X} &=& \Delta \phi\left(\vec p_{\rT,\mathrm{rec}},\vec
p_{\rT,X}\right)
\label{eq:phirecdef}
\eea
between the transverse momentum of the recoil, 
\bea
\vec p_{\rT,\mathrm{rec}} &=&
\sum_{i=t_1, t_2, b_1, b_2} \vec p_{\rT,i} \,,
\label{eq:recdef}
\eea
and the various objects $X$ of the \ttbb system, \ie the harder and softer top
quarks ($t_1, t_2$) and the harder and softer $b$-jets ($b_1,b_2$),
as well as the top-quark and the $b$-jet pairs.
These angular observables, sketched in \reffi{fig:recoilsketch}, 
reveal whether the respective object $X$ absorbs a significant fraction of the QCD recoil
through the presence (or absence) of peaks at $\Dphirec{X}=\pm \pi$.

\begin{figure}[p]
\tworecoilplots{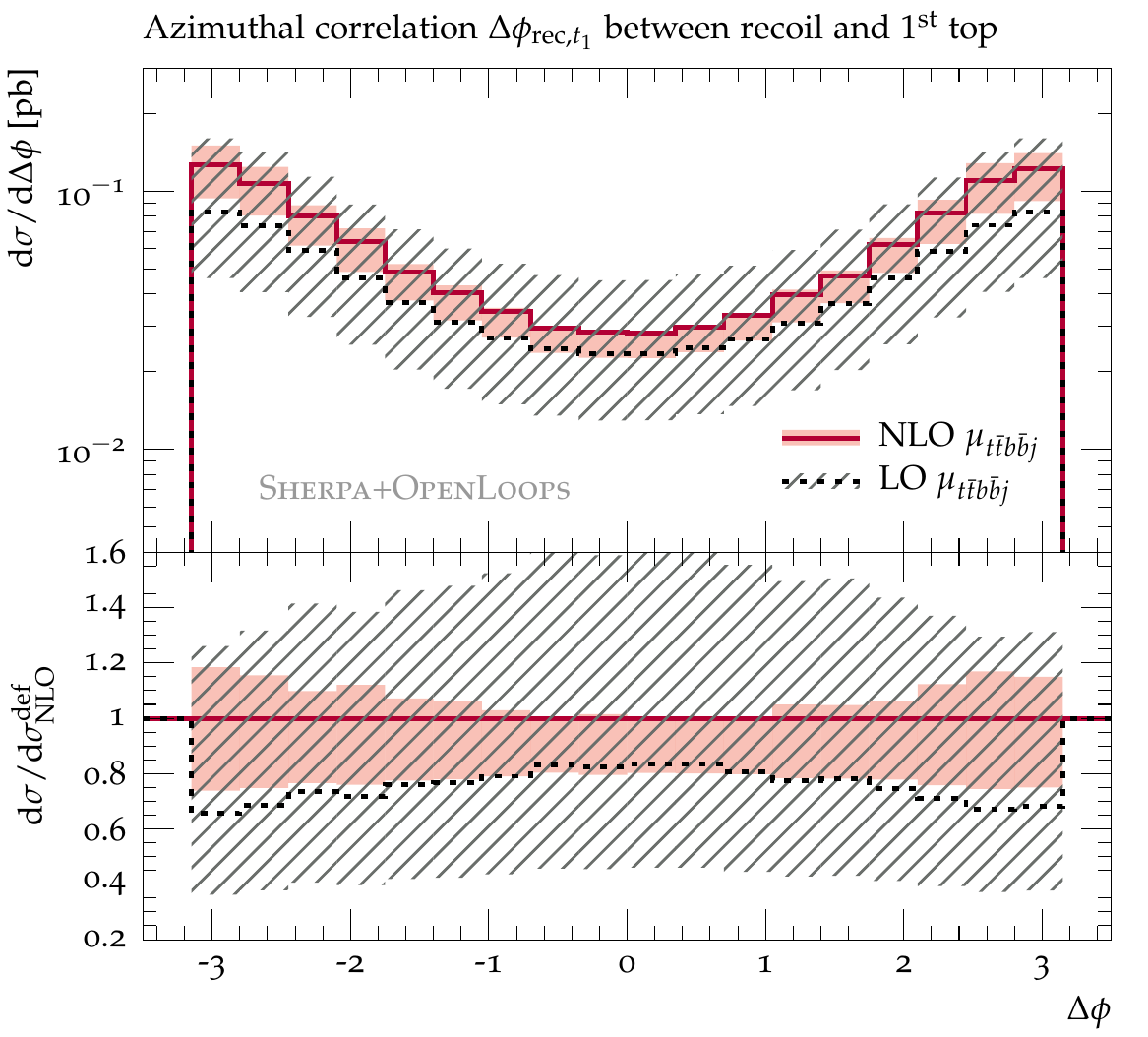}{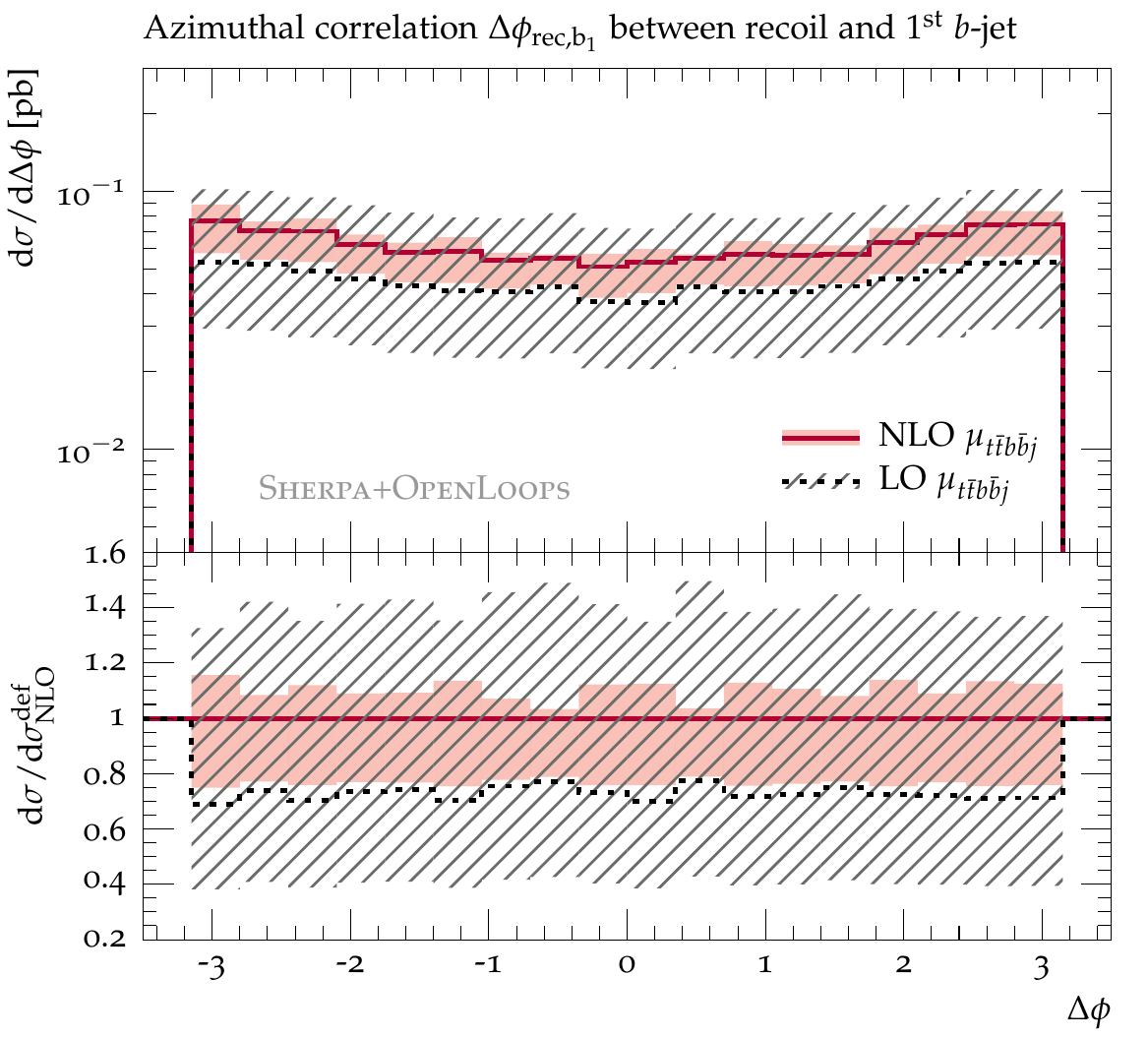}
\vspace*{-4ex}
\tworecoilplots{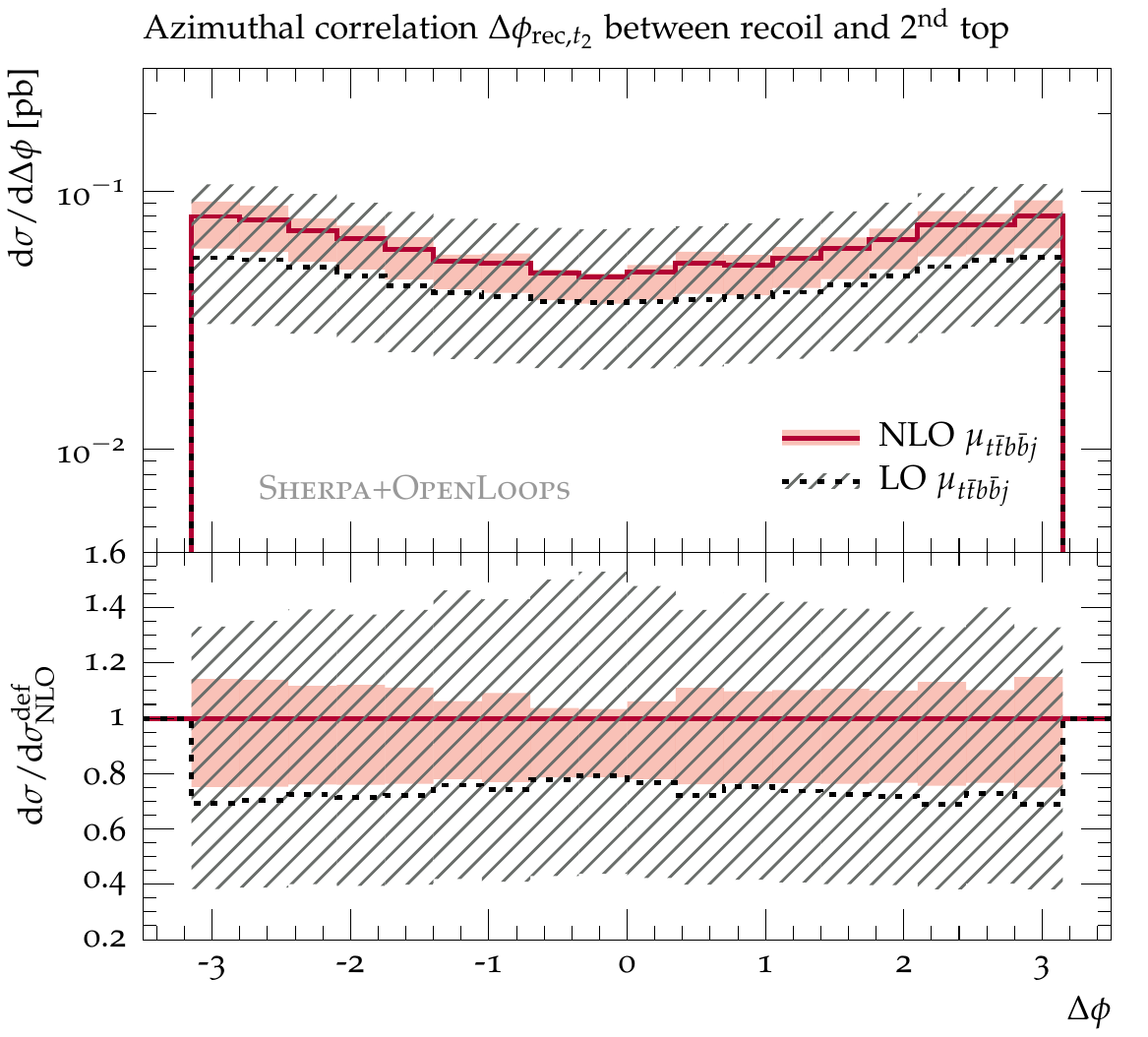}{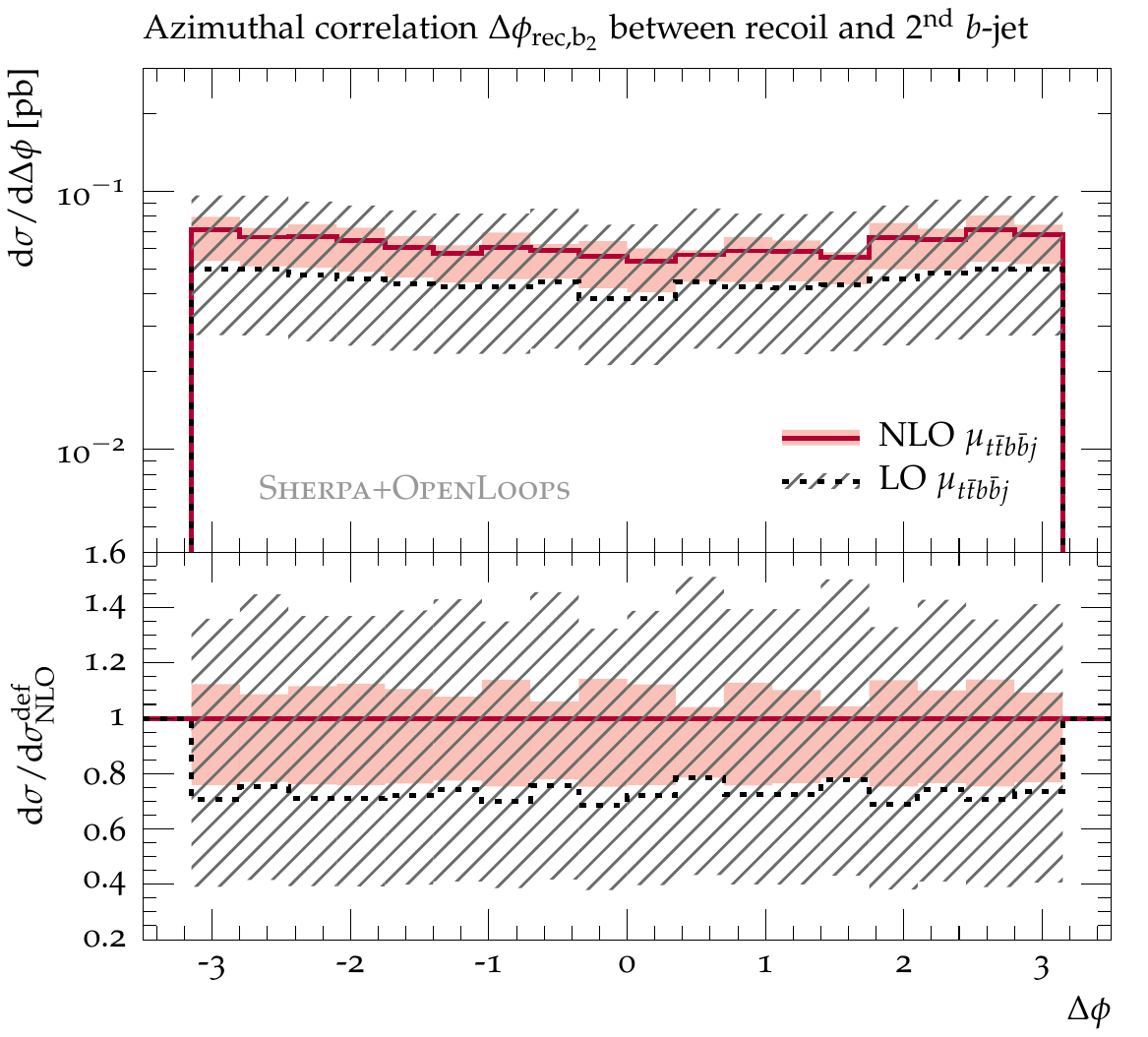}
\vspace*{-4ex}
\tworecoilplots{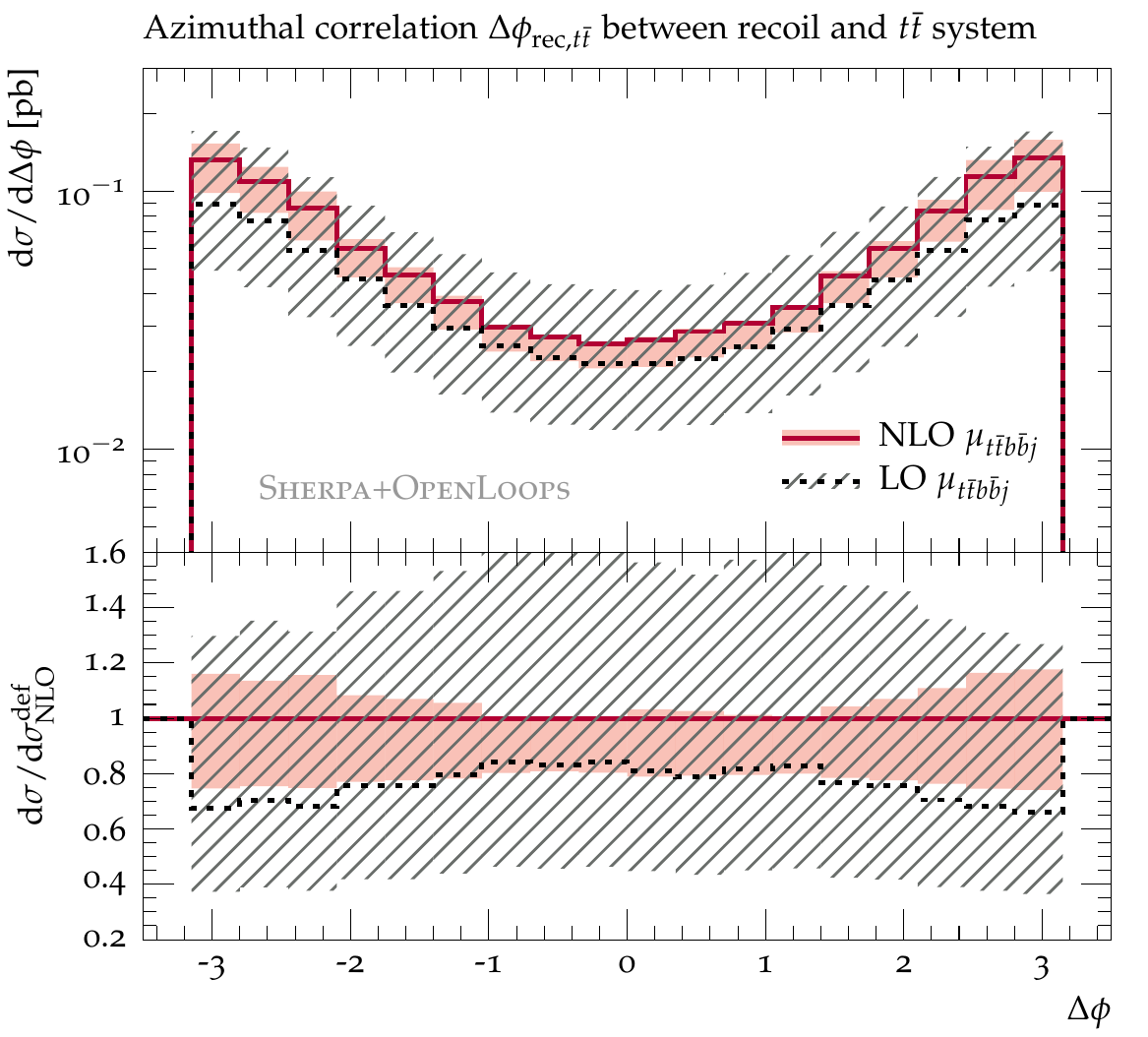}{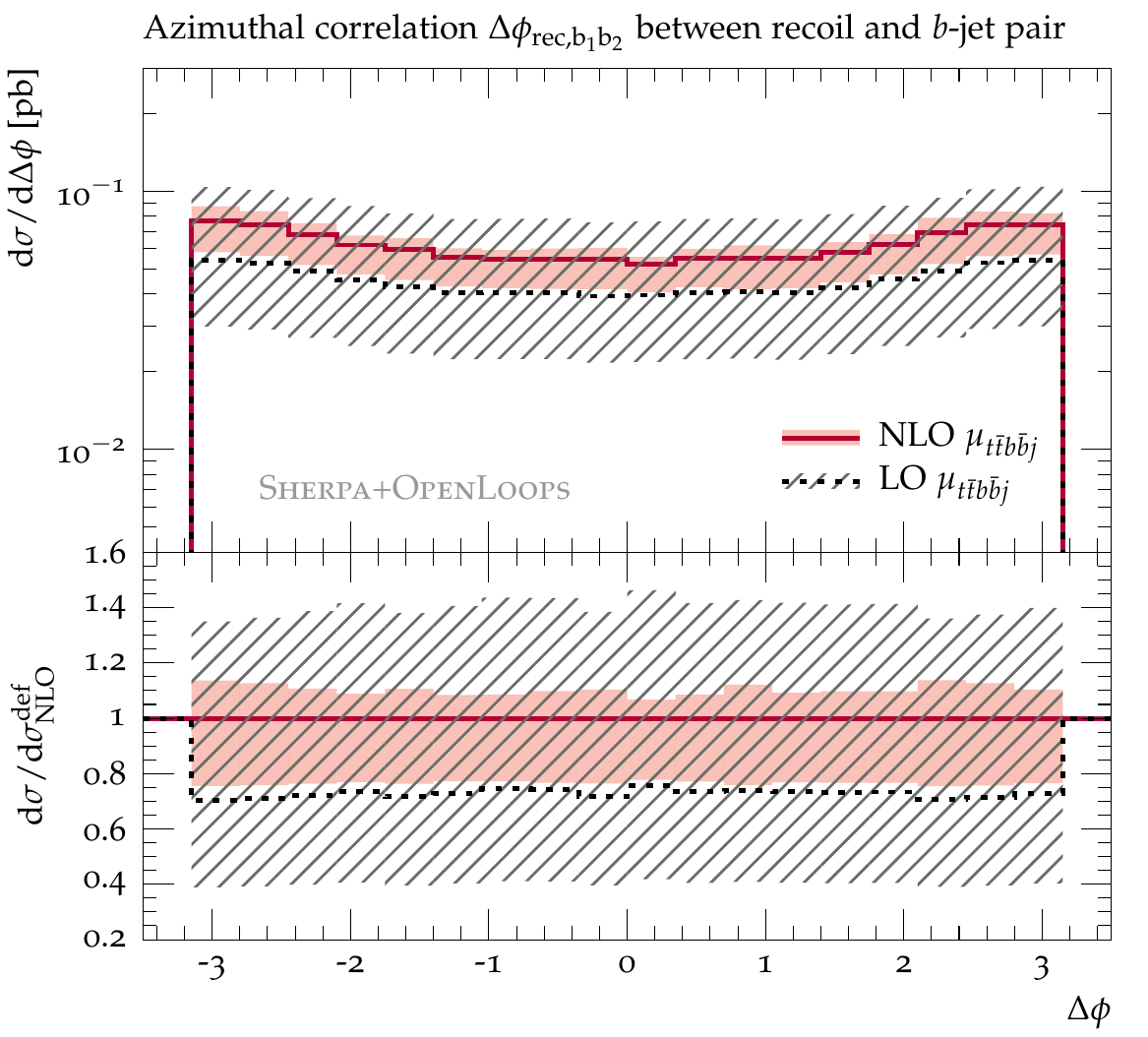}
\vspace*{-4ex}
\caption{
Distributions in the azimuthal angular separation 
$\Dphirec{X}$ between individual objects $X$ of the \ttbb system
and its recoil. 
See \refeq{eq:phirecdef}--\refeq{eq:recdef}.
The left column shows the angular correlations between the recoil and the
top-quark objects $X=t_1, t_2, t_1t_2$, where $t_1t_2$ denotes the top-pair
system.  
Corresponding observables for $b$-jet objects, $X=b_1, b_2, b_1 b_2$, are
shown in the right column.
Same setup and plots as in \reffi{fig:sud:2_logPT_J1_lowpt}.
}
\label{figs:recoilobs}
\end{figure}

In \reffi{figs:recoilobs} we present LO and NLO predictions for 
the azimuthal correlations between the recoil and the
various top-quark and $b$-jet objects.  
For these observables we focus on
the default scale, $\mur=\mudef$,  with seven-point variations.
The absolute distributions in the upper frames indicate a very clear
pattern: the recoil is preferentially absorbed by the harder top quark, 
and consequently also by the $t\bar t$ system, while the softer top quark and the
$b$-jets feature only weak angular correlations with respect to the recoil.
More precisely, in the case of the harder top, at $\Delta\phi=\pm\pi$ the
cross section is almost five times larger as compared to the central region,
while in the case of the harder\,(softer) $b$-jet this enhancement goes down
to about 50\% (20\%).
Thus it should be clear that naive shower models that distribute the recoil in a
democratic way may lead to a significant mismodelling of the $b$-jet
kinematics.
Concerning the accuracy of NLO predictions in \reffi{figs:recoilobs}, we
observe that all distributions are quite stable wrt to NLO corrections and scale
variations.
The most significant shape effects show up in the case of top-quark observables,
where scale uncertainties can shift the level of the recoil peak by
15--20\%, while for $b$-jets the flatness of the
azimuthal correlations is remarkably stable with respect to higher-order
effects.

These results demonstrate that fixed-order NLO predictions for $pp\to\ttbbj$
can be used as a precision benchmark to validate the modelling of
recoil effects in Monte Carlo simulations of \ttbb production.

\clearpage

\section{Tuning of QCD scale choice in $\boldsymbol{\ttbb}$ production}
\label{se:ttbb_tuning}

In the literature on $pp\to \ttbb$ at NLO, the usage of dynamic scales of
type \mbox{$\mur=\muttbb$~\refeq{eq:muttbb}} has been advocated on the basis of the
moderate size of the resulting NLO correction factor, $K=\sigmaNLO/\sigmaLO$.
However, as pointed out in~\cite{Jezo:2018yaf}, the smallness of the
observed $K$-factor was largely due to the usage of a rather high LO value of
$\alpha_{\rS}$ as input for $\sigma_\LO$, while using the same $\alphaS$ in
$\sigmaLO$ and $\sigmaNLO$ results in a correction factor as large as $K\simeq
1.9$~\cite{Jezo:2018yaf}.
The lack of perturbative convergence, reflected by this large $K$-factor,
may simply be the consequence of the fact that $\mur=\muttbb$ is a
suboptimal choice.
At the same time, it may also be the origin of the discrepancies between NLOPS simulations of \ttbb
production~\cite{talkPozzoriniTOP2018}.
In fact, when matrix elements at NLO are matched to parton showers, the
spectrum of the hardest QCD emission receives uncontrolled corrections of
order $(K-1)=\ord(\alphaS)$.  Such effects are formally beyond NLO, but for
$K\gg 1$ they can lead to sizeable distortions of the radiation
spectrum~\cite{talkPozzoriniTOP2018}.

In the light of these observations, and given the strong scale dependence of
the \ttbb \mbox{$K$-factor,} it is clear that a relatively mild reduction of 
the nominal scale would automatically lead to a smaller $K$-factor 
and, possibly, also to an improved behaviour of NLO matched 
simulations.
However, the large \ttbb \mbox{$K$-factor} may also be due to 
large higher-order effects that are not related to the choice 
of $\mur$. In this case, a reduction of the $K$-factor 
via $\mur$ rescaling would only give a misleading impression of 
perturbative convergence without curing any problem.
These considerations raise the question whether a reduction of the
\ttbb \mbox{$K$-factor} through a smaller choice of $\mur$ may be supported 
through solid theoretical arguments.
Generic considerations based on naturalness and perturbative
convergence point towards a reduction of the standard 
\ttbb scale choice by a factor $1/2$ to $1/3$~\cite{talkPozzoriniTOP2018}.
However, only the knowledge of the next perturbative order
can shed full light on the goodness of a
scale choice, \ie on its effectiveness in capturing the dominant
higher-order effects.
In the case at hand, the \ttbb scale choice could be tuned
based on the requirement that 
\bea 
\sigma^{\ttbb}_{\NLO}(\muropt,\mufopt)&\mbeq&\sigma^{\ttbb}_\NNLO(\mur,\muf)\,, 
\label{eq:NNLOincmatching}
\eea
\ie by optimising the choice of the scales $\murfopt$ in such a way 
that NLO \ttbb predictions match NNLO ones.\footnote{The
reference scales $\mu_{\rR,\rF}$ used at NNLO can be chosen and varied in
different ways. However, due the small level of expected scale dependence at
NNLO, such choices should not have a dramatic impact on the tuned scales
$\murfopt$. Note also that equation~\refeq{eq:NNLOincmatching} may have no
exact solution, in which case it should be understood as the requirement 
of a minimal difference between the NLO and NNLO cross sections.}
However, the required NNLO calculation is completely out of reach.
Nonetheless, the NLO corrections to $pp\to \ttbbj$ presented in this paper
represent one of the building blocks of \ttbb production at NNLO,
and as such they can provide useful insights on how to improve the
\ttbb scale choice.
The idea is that the condition \refeq{eq:NNLOincmatching}
can be imposed at the level of the jet-radiation spectrum 
by requiring
\bea 
\frac{\rd\sigma^{\ttbb}_{\NLO}}{\rd \pTj}(\muropt,\mufopt)&\mbeq&
\frac{\rd\sigma^{\ttbb}_\NNLO}{\rd\pTj}(\mur,\muf)\,=\,
\frac{\rd\sigma^{\ttbbj}_\NLO}{\rd\pTj}(\mur,\muf)\,.
\label{eq:NNLOptmatching}
\eea
With other words, the scale choice can be 
tuned in such a way that the tree-level description of 
the jet-$\pT$ spectrum that results from the \ttbb NLO
calculation matches the more precise prediction of the \ttbbj NLO
calculation.
Contrary to \refeq{eq:NNLOincmatching}, this procedure cannot guarantee the
correct description of higher-order effects at the level of the inclusive 
\ttbb cross section.\footnote{We note that this approach does not improve
the precision of the integrated \ttbb cross sections. Its goal is only to
optimise the choice of the central scale.
} 
Nonetheless it is attractive for at least two reasons.
First, tuned \ttbb NLO predictions will guarantee a much more accurate
description of the jet-$\pT$  spectrum, which is known to play a critical
role in Monte Carlo simulations.
Second, the shape of the jet-$\pT$ spectrum can be used to judge the quality of
the matching procedure~\refeq{eq:NNLOptmatching}, and the
general consistency of the procedure can be validated 
by comparing various other jet observables.

The results of this tuning procedure are presented in \reffi{fig:tuningPTJ},
where we show the distribution in the $\pT$ of the hardest light jet, and in the
invariant masses of the systems formed by the hardest light jet in combination
with the leading or the subleading $b$-jet.
The tuning is carried out through a constant rescaling of the standard \ttbb
scale choice,
\bea
(\mur,\muf)&=&(\kappa\,\muttbb,\kappa\,\frac{\HT}{2})\,,
\label{eq:tuningtransf}
\eea
such as to match NLO predictions for the integrated \regionttbbj  
cross section based on the default scale $\mudef$.
To be conservative, we have compared two possible ways of tuning the \ttbb
scale.  
In the first approach, the rescaled \ttbb NLO predictions are matched to 
nominal \ttbbj NLO predictions,
whereas in the second approach the tuning is done by matching 
the average values of the respective seven-point variation
bands.
The outcome of these two matching prescriptions 
is shown in the left
and right columns of \reffi{fig:tuningPTJ}.
Matching nominal predictions leads to a reduction of the 
default \ttbb scale by a factor%
\footnote{We have checked that keeping $\muf=\HT/2$ fixed and tuning only
$\mur$ would require a rescaling factor $\kappa=1/1.76$.}
$\kappa=1/1.6$, whereas matching the scale-variation bands in a symmetric way
requires a significantly smaller rescaling, $\kappa=1/1.14$.  
This large difference is mainly due to the strong asymmetry of the factor-two
variation band of the tree-level prediction, \ie $pp\to \ttbb$ at NLO.
In this respect, we note that such asymmetry is mainly due to the
logarithmic nature of the scale dependence~\refeq{eq:LOmurdep}. Thus the 
asymmetry of the LO band would largely disappear on logarithmic scale, and
the prescriptions based on the central scale and the average of the bands
would be significantly closer to each other.

For all considered jet observables we find that both tuning scenarios lead
to a very good agreement, not only in the normalisation, but also at the
level of shapes.
The findings of this analysis support a reduction of the standard \ttbb
scale \refeq{eq:tuningtransf} by up to a factor $\kappa \sim 1/1.6$. In the
\regionttbb\!(\regionttb) phase space, $\kappa = 1/1.6$
corresponds to a reduction of the \ttbb $K$-factor from 
1.80\,(1.92) to 1.51\,(1.62) and an increase of the nominal 
\ttbb cross section by 18\% (21\%). 

\begin{figure}[t]
\bce\twotuningplots{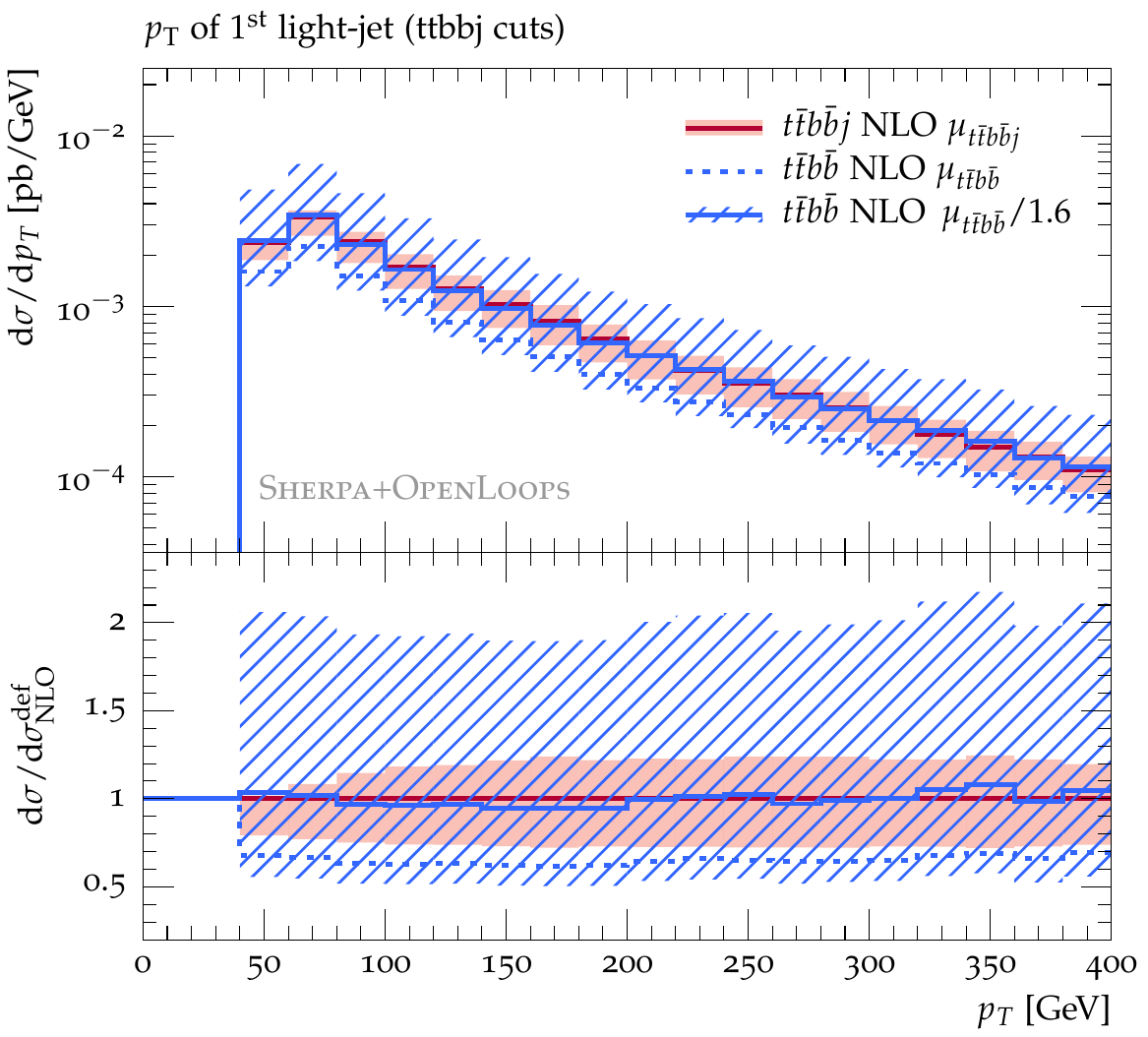}{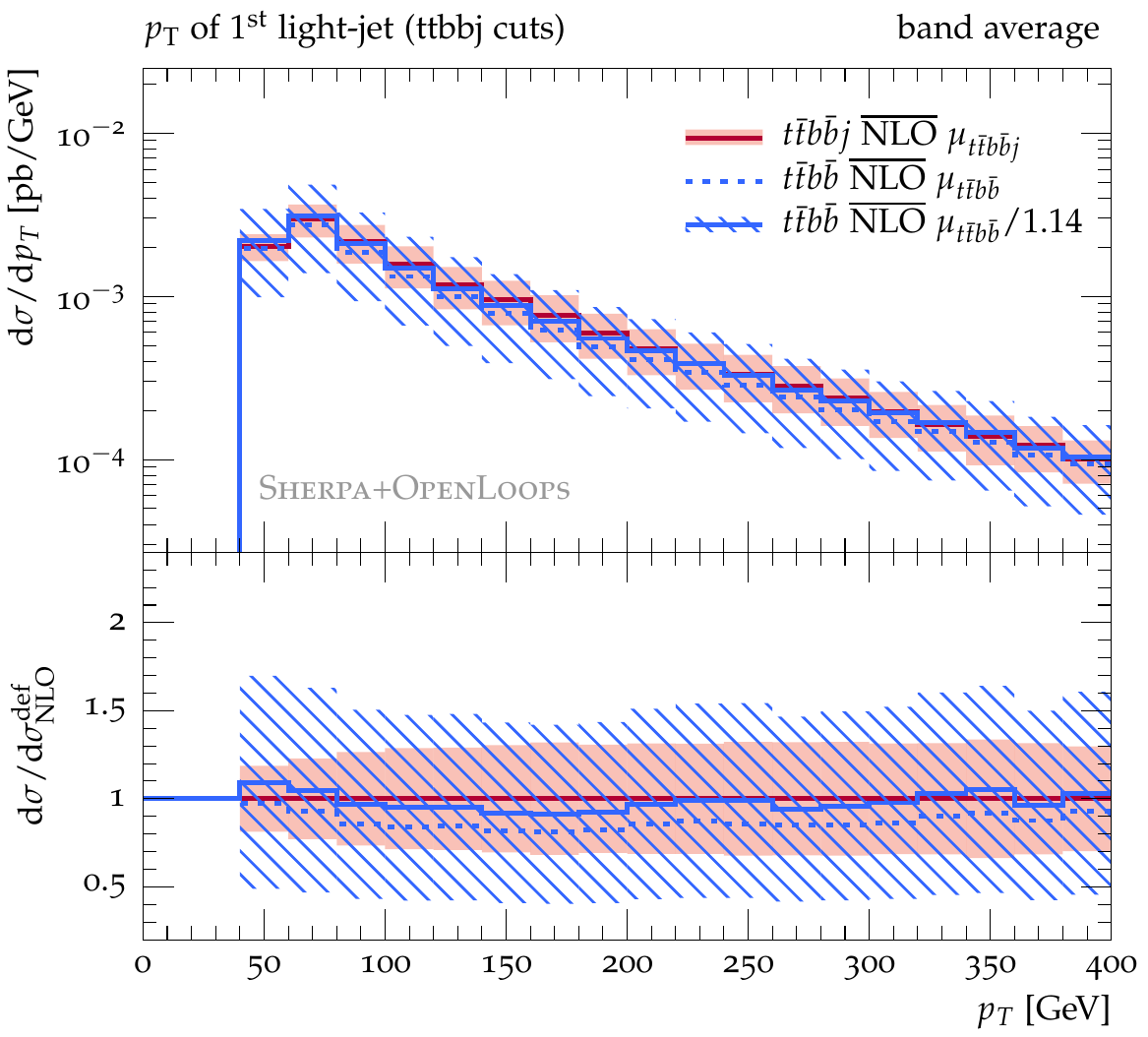}\ece
\vspace*{-4ex}
\bce\twotuningplots{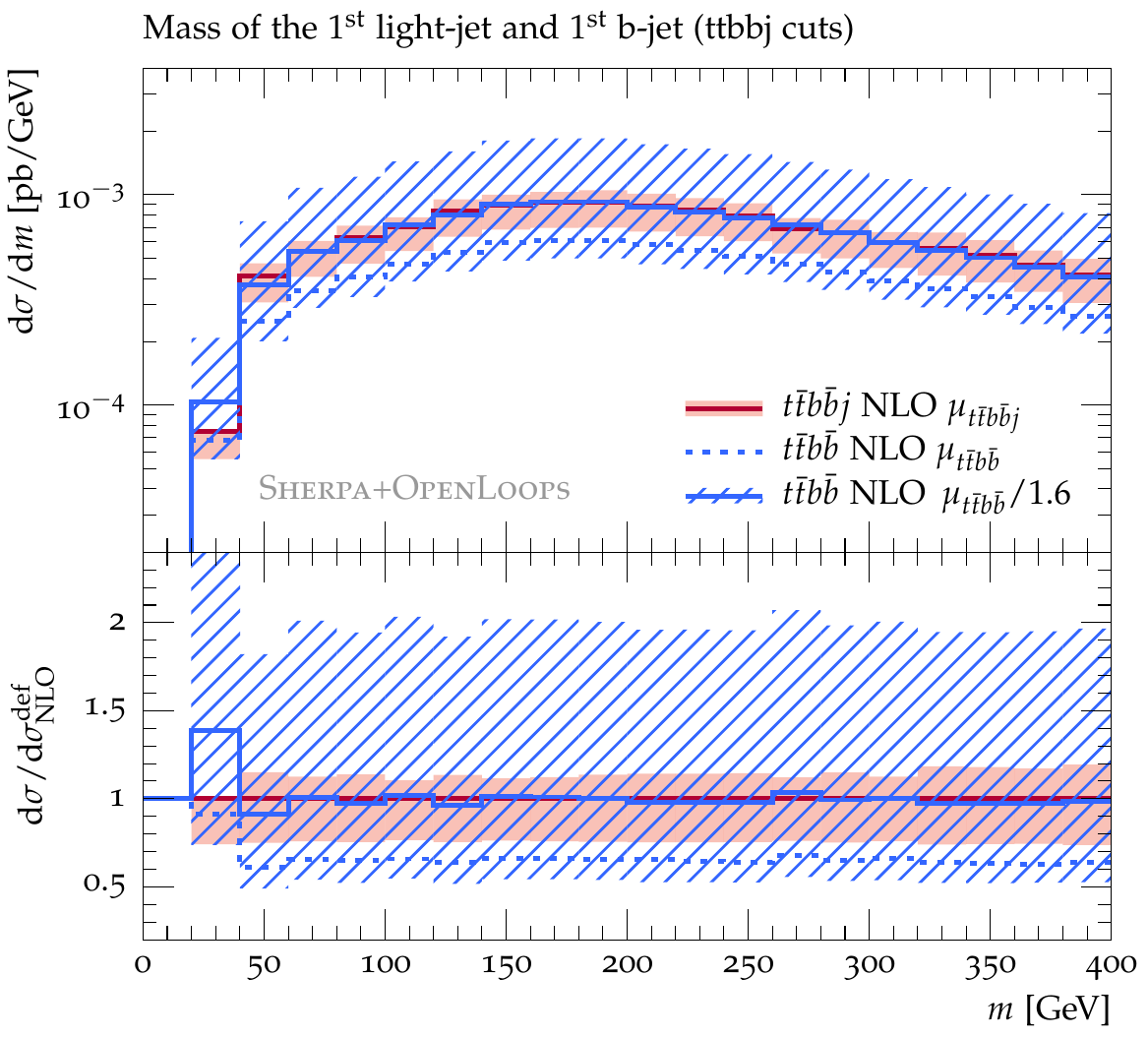}{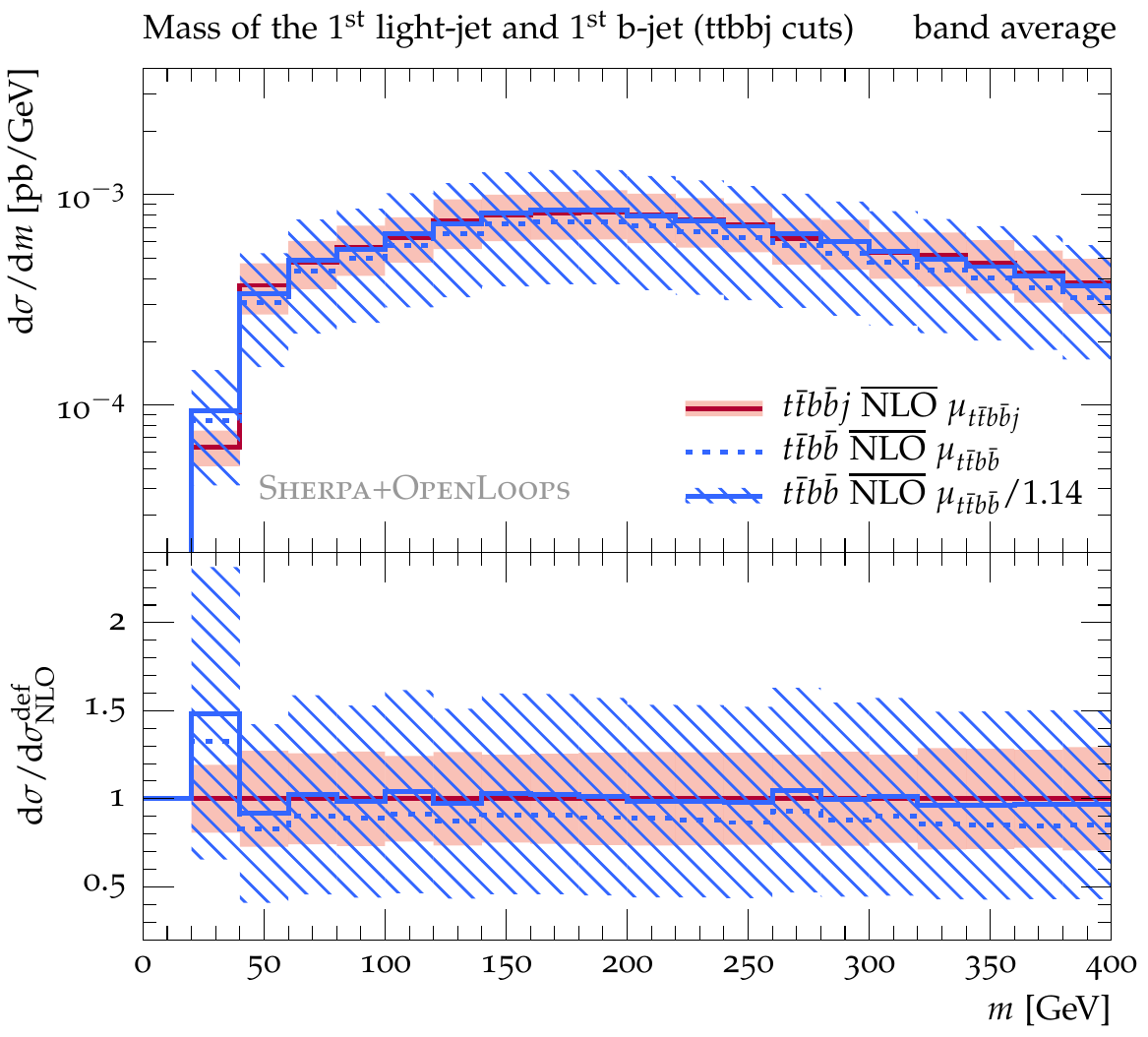}\ece
\vspace*{-4ex}
\bce\twotuningplots{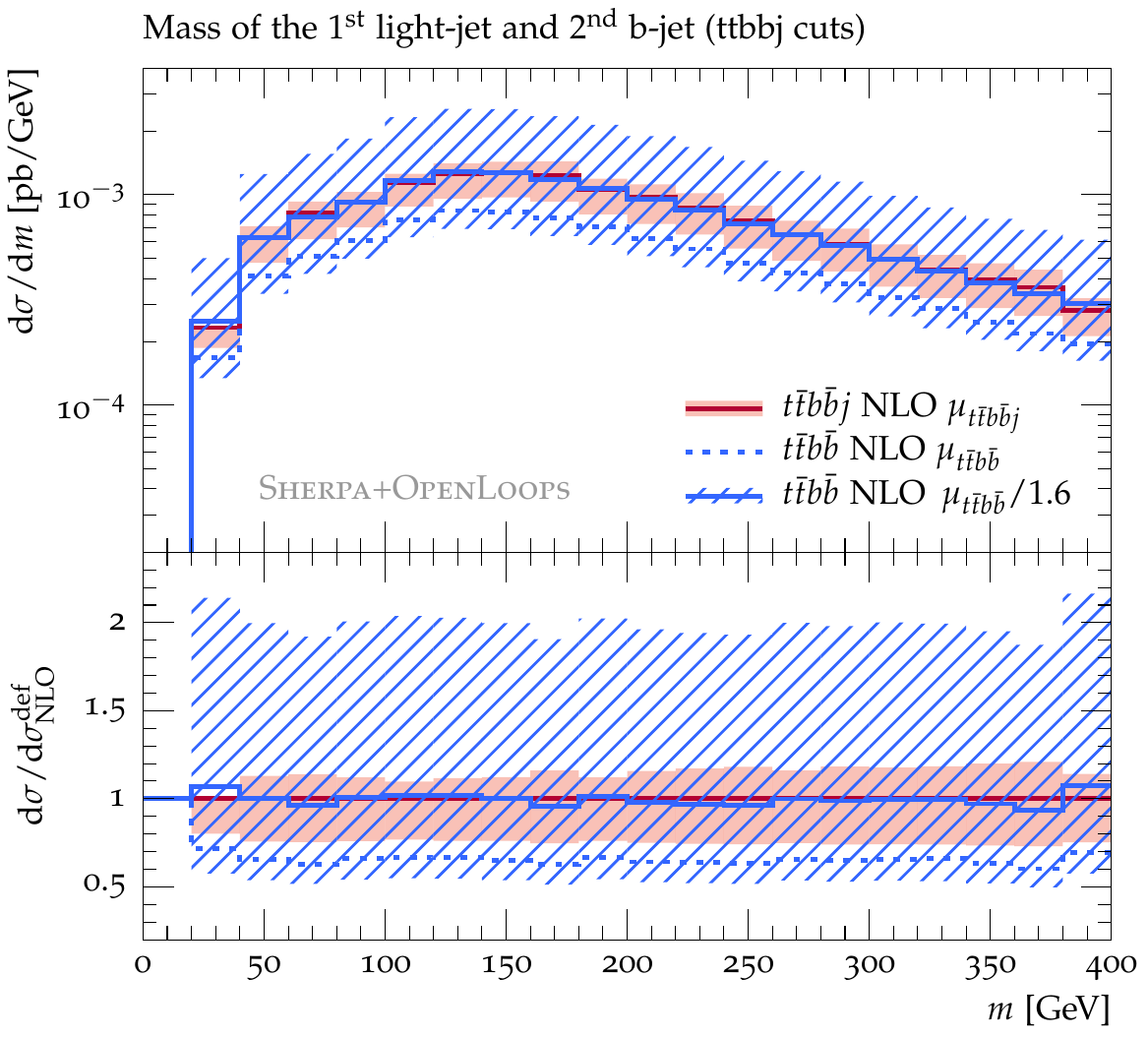}{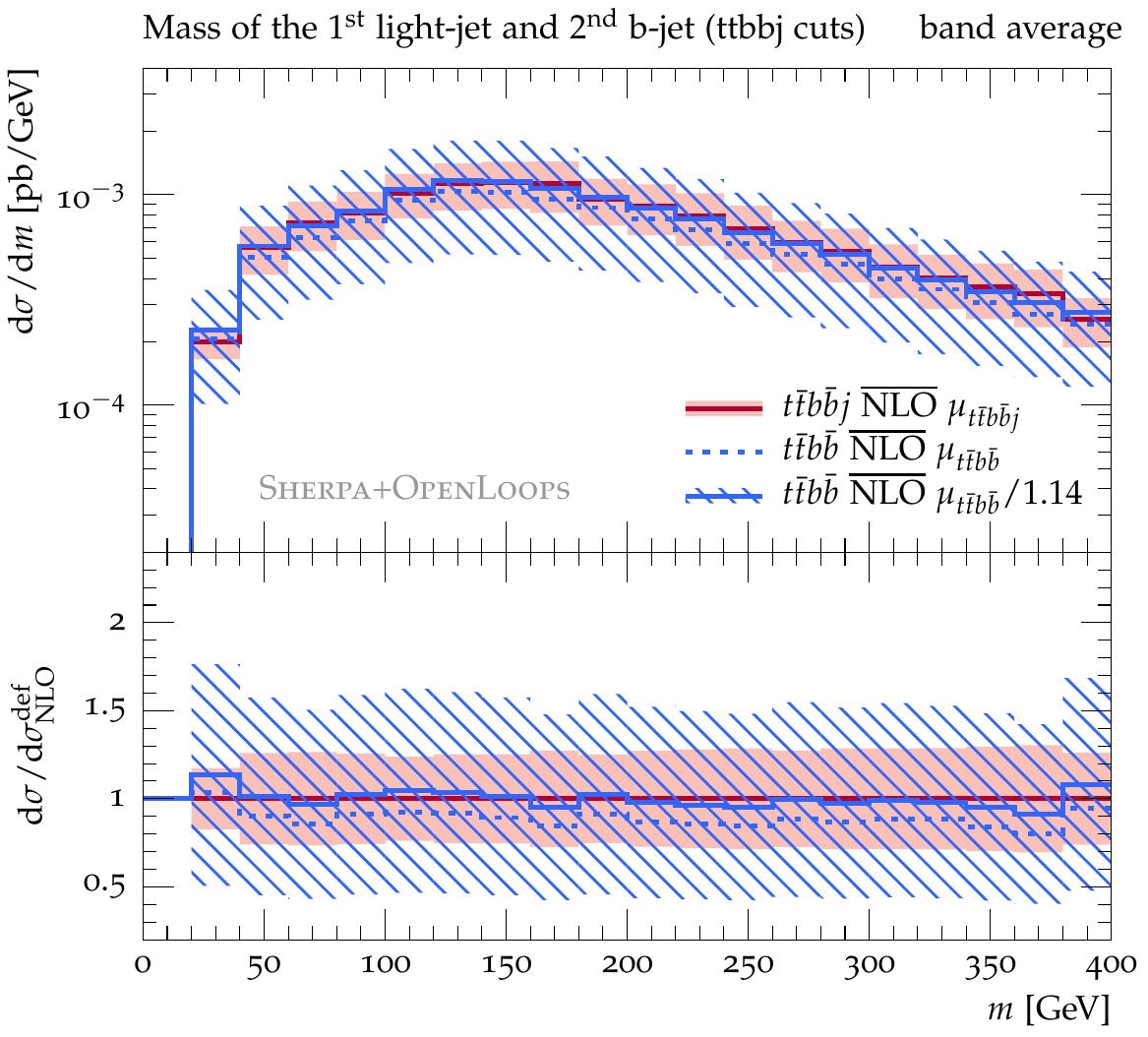}\ece
\vspace*{-4ex}
\caption{
Distributions in the $\pT$ of the leading jet and 
the mass of light-jet--$b$-jet systems
in the \regionttbbj phase space.  
Comparison of NLO \ttbbj predictions with default scale choice,
$(\mur,\muf)=(\mudef,\HT/2)$, to NLO \ttbb predictions with
$(\mur,\muf)=(\kappa\,\muttbb,\kappa\,\HT/2)$.
In the left plots, the reference curves for the matching procedure 
(solid, labelled NLO) correspond to the above central scales, and the applied 
rescaling factor is $\kappa=1/1.6$.  
In the right plots, the reference curves (solid, labelled $\overline{\NLO}$)
are the average values of the scale-variation bands, and $\kappa=1/1.14$.
The blue dashed curves indicate the position of the NLO \ttbb reference
prediction before tuning, while all other NLO \ttbb predictions and
scale-variation bands correspond to the tuned scales.
}
\label{fig:tuningPTJ}
\end{figure}

\clearpage

\section{Summary}
\label{se:conclusions}

Measurements of $\ttbar H(\bbbar)$ production at the LHC require very
accurate theoretical simulations of the irreducible \ttbb background.
To address the dominant sources of systematic uncertainties, which stem from the
modelling of QCD radiation in \ttbb events, we have presented a
calculation of \ttbb production in association with one extra jet at NLO
QCD.

To carry out this non-trivial calculation we used 
\OLtwo  in combination
with the {\Sherpa} and {\sc Munich} Monte Carlo frameworks.
Technically, the calculation of the required $2\to 5$ one-loop 
amplitudes has confirmed that 
the new algorithms implemented in \OLtwo can tackle multi-particle and 
multi-scale problems with very high CPU efficiency and 
numerical stability.

We have discussed $pp\to \ttbbj$ at the
13\,TeV LHC with emphasis on the effects of NLO corrections and scale uncertainties. 
To this end, we have studied conventional factor-two rescalings, as well as
variations of the kinematic dependence of dynamic scales.  In order to
disentangle normalisation and shape uncertainties in a transparent way, we
have proposed to compare dynamic scales upon alignment of the NLO maxima of
the respective scale-variation curves.

In general, the typical level of scale uncertainties in $pp\to\ttbbj$ at NLO
is \percentrange{20}{30} for integrated cross sections and below 10\% in the shapes of
distributions.
The calculation at hand can thus be used as a precision benchmark to validate 
the modelling of QCD radiation in Monte Carlo generators of \ttbb production.
With this motivation in mind, we have presented NLO predictions for various 
azimuthal correlations that provide a transparent picture of the
effects of the recoil of QCD radiation on the different objects of the
\ttbb system.

Finally, we have discussed the issue of the large NLO $K$-factor observed in
inclusive NLO calculations of \ttbb production, and we have addressed the
question of whether it is justified to reduce this $K$-factor through ad-hoc
scale choices.
In this respect we have argued that the 
NLO corrections to $pp\to \ttbbj$ entail information on
$pp\to \ttbb$ beyond NLO, which can be exploited 
to identify an optimised scale choice. 
Specifically, we have proposed the idea of adjusting the nominal \ttbb scale choice
such as to match the jet emission rate 
predicted by $pp\to\ttbbj$ at NLO. 
This improved scale choice leads to a reduction of the \ttbb $K$-factor, and is also expected to 
attenuate theoretical uncertainties in the context of NLO matching to parton showers.

\vspace*{-1ex}
\acknowledgments

The motivation for this project stems from Monte Carlo generator studies
carried out within the LHC Higgs Cross Section Working group.  In this
respect we are especially grateful to Tom\'a\v{s} Je\v{z}o, Frank Siegert and Marco
Zaro for numerous and valuable discussions.
We gratefully acknowledge extensive technical support from Stefan H\"oche,
Tom\'a\v{s} Je\v{z}o, Jean-Nicolas Lang, Jonas Lindert and Hantian Zhang.
This research was supported by the Swiss National Science
Foundation~(SNF) under contract BSCGI0-157722.
MZ acknowledges support by the Swiss National Science
Foundation (Ambizione grant PZ00P2-179877).
The work of SK is supported by the ERC Starting Grant 714788 REINVENT. 

\bibliographystyle{JHEP}
\bibliography{ttbbj}

\end{document}

%% file: texmacros.tex
\newcommand{\QCD}{\mathrm{QCD}}

\newcommand{\dyn}{\mathrm{dyn}}
\newcommand{\cut}{\mathrm{cut}}

\newcommand{\LO}{\mathrm{LO}}

\newcommand{\NLO}{\mathrm{NLO}}
\newcommand{\NNLO}{\mathrm{NNLO}}

\newcommand{\GeV}{\text{GeV}\xspace}

\newcommand{\rd}{\mathrm d}

\newcommand{\rT}{\mathrm{T}}
\newcommand{\rF}{\mathrm{F}}
\newcommand{\rR}{\mathrm{R}}
\newcommand{\rS}{\mathrm{S}}

\newcommand{\ord}{\mathcal{O}}

\renewcommand{\it}[1]{\textit{#1}}

\newcommand{\ie}{i.e.~}
\newcommand{\eg}{e.g.~}

\newcommand{\as}{\alpha_\mathrm{S}}
\newcommand{\alphaS}{\as}
\newcommand\asfour{\alpha^{(4\rF)}_{\rm S}}
\newcommand\asfive{\alpha^{(5\rF)}_{\rm S}}

\newcommand\murfopt{\ensuremath{\mu^{\mathrm{opt}}_{\rR,\rF}}}
\newcommand\mufopt{\ensuremath{\mu^{\mathrm{opt}}_\rF}}
\newcommand\muropt{\ensuremath{\mu^{\mathrm{opt}}_\rR}}
\newcommand\muf{\ensuremath{\mu_\rF}}
\newcommand\mur{\ensuremath{\mu_\rR}}
\newcommand\xir{\ensuremath{\xi_\rR}}
\newcommand\xif{\ensuremath{\xi_\rF}}
\newcommand{\muF}{\muf}
\newcommand{\muR}{\mur}
\newcommand{\HT}{\ensuremath{H_\rT}}

\newcommand{\scalettbb}[1]{\ensuremath{{#1}_{\ttbb}}}
\newcommand{\scalettbbj}[1]{\ensuremath{{#1}_{\ttbbj}}}
\newcommand{\scaledef}[1]{\scalettbbj{#1}}
\newcommand{\scalembb}[1]{\ensuremath{{#1}_{g\bbbar}}}
\newcommand{\scalehtbbj}[1]{\ensuremath{{#1}_{\rT, \mathrm{jets}}}}
\newcommand{\scaleht}[1]{\ensuremath{{#1}_{\rT, \mathrm{tot}}}}

\def\htjets{H_{\rT,\mathrm{jets}}}

\newcommand\Dphirec[1]{\Delta\phi_{\mathrm{rec},X}}
\newcommand{\ptav}[1]{\langle p_{\rT,#1}\rangle}
\newcommand{\ptratav}[1]{\langle p_{\rT,j_1}/p_{\rT,#1}\rangle}

\newcommand{\mumax}{\mu_\mathrm{max}}
\newcommand{\mudyn}{\mu_\dyn}
\newcommand{\barmudyn}{\bar\mu_\dyn}
\newcommand{\mudyni}[1]{\mu_{\dyn,#1}}
\newcommand{\tildemudyni}[1]{{\tilde \mu}_{\dyn,#1}}
\newcommand{\barmudyni}[1]{\bar\mu_{\dyn,#1}}

\newcommand{\muttbb}{\scalettbb{\mu}}
\newcommand{\muttbbj}{\scalettbbj{\mu}}
\newcommand{\mudef}{\scaledef{\mu}}
\newcommand{\mumbb}{\scalembb{\mu}}
\newcommand{\muhtbbj}{\scalehtbbj{\mu}}
\newcommand{\muht}{\scaleht{\mu}}

\newcommand{\hatmudef}{\scaledef{\tilde\mu}}
\newcommand{\hatmumbb}{\scalembb{\tilde\mu}}
\newcommand{\hatmuhtbbj}{\scalehtbbj{\tilde\mu}}
\newcommand{\hatmuht}{\scaleht{\tilde\mu}}

\newcommand{\sigmaLO}{\ensuremath{\sigma_{\mathrm{LO}}}}
\newcommand{\sigmaNLO}{\ensuremath{\sigma_{\mathrm{NLO}}}}
\newcommand{\sigmaLOdef}{\ensuremath{\sigma_{\mathrm{LO,def}}}}
\newcommand{\sigmaNLOdef}{\ensuremath{\sigma_{\mathrm{NLO,def}}}}

\newcommand{\FastJet}{{\rmfamily\scshape FastJet}\xspace}
\newcommand{\Collier}{{\rmfamily\scshape Collier}\xspace}
\newcommand{\Sherpa}{{\rmfamily\scshape Sherpa}\xspace}
\newcommand{\Comix}{{\rmfamily \scshape Comix}\xspace}
\newcommand{\SherpaOpenLoops}{{\rmfamily\scshape Sherpa+OpenLoops}\xspace}

\newcommand{\MunichOpenLoops}{{\rmfamily \scshape Munich+OpenLoops}\xspace}

\newcommand{\Munich}{{\rmfamily \scshape Munich}\xspace}
\newcommand{\Matrix}{{\rmfamily \scshape Matrix}\xspace}
\newcommand{\OpenLoops}{{\rmfamily\scshape OpenLoops}\xspace}
\newcommand{\OLL}{\mbox{\OpenLoops\hspace{-0.5mm}2}\xspace}
\newcommand{\OLtwo}{\OLL}

\newcommand{\OL}{\OpenLoops}

\newcommand{\Rivet}{{\rmfamily\scshape Rivet}\xspace}
\newcommand{\Recola}{{\rmfamily\scshape Recola}\xspace}

\newcommand{\ttbar}{\ensuremath{t \bar t}\xspace}
\newcommand{\bbbar}{\ensuremath{b \bar b}\xspace}
\newcommand{\ttbb}{\ensuremath{t \bar t b\bar b}\xspace}
\newcommand{\ttbbj}{\ensuremath{t\bar{t}b\bar{b}j}\xspace}

\newcommand{\ttbbjets}{\ensuremath{t\bar{t}b\bar{b}+}jets\xspace}

\newcommand{\regionttb}{\texttt{ttb}\xspace}
\newcommand{\regionttbb}{\texttt{ttbb}\xspace}
\newcommand{\regionttbj}{\texttt{ttbj}\xspace}
\newcommand{\regionttbbj}{\texttt{ttbbj}\xspace}
\newcommand{\regionttbjj}{\texttt{ttbjj}\xspace}
\newcommand{\regionttbbjj}{\texttt{ttbbjj}\xspace}

\newcommand{\pT}{p_{\rT}}
\newcommand{\Nb}{N_b}
\newcommand{\Nj}{N_j}
\newcommand{\Nbmin}{N^{\mathrm{min}}_b}
\newcommand{\Njmin}{N^{\mathrm{min}}_j}

\newcommand\pTj{\ensuremath{p_{\rT,j}}}
\newcommand\pTb{\ensuremath{p_{\rT,b}}}

\newcommand\pTcut{\ensuremath{p^{\cut}_{\rT}}}

\def\DRbb{\Delta R_{bb}}

\newcommand{\mbeq}{\overset{!}{=}}

\def\bit{\begin{itemize}}
\def\eit{\end{itemize}}

\def\be{\begin{equation}}
\def\ee{\end{equation}}

\def\bea{\begin{eqnarray}}
\def\beqar{\bea}
\def\eea{\end{eqnarray}}
\def\eeqar{\eea}

\def\bce{\begin{center}}
\def\ece{\end{center}}

\def\refeq#1{\mbox{(\ref{#1})}}

\def\reffi#1{\mbox{Fig.~\ref{#1}}}
\def\reffis#1#2{\mbox{Figures~\ref{#1}--\ref{#2}}}
\def\refta#1{\mbox{Table~\ref{#1}}}

\def\reftastwo#1#2{\mbox{Tables~\ref{#1} and \ref{#2}}}
\def\refse#1{\mbox{Section~\ref{#1}}}
\def\refses#1#2{\mbox{Sections~\ref{#1}--\ref{#2}}}

\def\percentrange#1#2{\mbox{$#1$--$#2$\%}}

\def\plotboxrescale{1.75}

\def\plots{plots}

\newcommand{\getmainP}[3]{%
\includegraphics[scale=0.33,trim=0 25 3 0,clip]{\plots/#1/#2/#3.pdf}
}

\newcommand{\getMratio}[3]{%
\includegraphics[scale=0.33,trim=0 25 3 15,clip]{\plots/#1/#2/#3.pdf}
}

\newcommand{\getBratio}[3]{%
\includegraphics[scale=0.33,trim=0 0 3 15,clip]{\plots/#1/#2/#3.pdf}
}

\newcommand{\makeplotA}[5]{
  \begin{minipage}{0.25\textwidth}
  \getmainP{#1}{#2}{#5}\\
  \getMratio{#1}{#4}{#5}\\
  \getBratio{#1}{#3}{#5} 
  \end{minipage}
}

\newcommand{\makeplotB}[5]{
  \begin{minipage}{0.25\textwidth}
  \getmainP{#1}{#2}{#5}\\
  \getMratio{#1}{#4}{#5}\\
  \getBratio{#1}{#3}{#5}  
  \end{minipage}
}

\newcommand{\plotobs}[1]{
\begin{tabular}{cc}
\scalebox{\plotboxrescale}{\makeplotA{ttbbj}{def_abs}{ht5_corr}{mbb_corr}{#1}} &
\scalebox{\plotboxrescale}{\makeplotB{ttbbj}{def_norm}{nlo_norm_uncorr}{lo_norm_uncorr}{#1}}
\end{tabular}
\vspace*{-4ex}
}

\newcommand{\showscanplots}[1]{\plotobs{#1}}
\newcommand{\scanplots}[1]{\showscanplots{#1}}

\newcommand{\plotnlodiag}[1]{
\begin{minipage}{0.3\textwidth}
\bce
\includegraphics[width=0.9\textwidth]{diagrams/#1}
\ece
\end{minipage}
}

\newcommand{\tworecoilplots}[2]{
\bce
\includegraphics[scale=0.6,trim=0 0 0 0,clip]{\plots/ttbbj/recoil_abs/#1}
\hspace{5mm}
\includegraphics[scale=0.6,trim=0 0 0 0,clip]{\plots/ttbbj/recoil_abs/#2}
\ece
}

\newcommand{\twotuningplots}[2]{
\bce
\includegraphics[scale=0.56,trim=0 0 0 0,clip]{\plots/ttbbj/tuning_plot_50GeV/#1}
\hspace{5mm}
\includegraphics[scale=0.56,trim=0 0 0 0,clip]{\plots/ttbbj/tuning_plot_50GeV/#2}
\ece
}
